\documentclass[reprint,aps,physrev,twocolumn,showpacs,showkeys, 10pt, longbibliography, floatfix, amsmath,amssymb, nolinenumbers
]{revtex4-2}

\usepackage{graphicx}% Include figure files
\usepackage{dcolumn}% Align table columns on decimal point
\usepackage{bm}% bold math
\usepackage{amsfonts}
\usepackage[colorlinks = true,linkcolor = blue,urlcolor  = blue,citecolor = blue,anchorcolor = blue]{hyperref}
\usepackage{float}
\usepackage{times}
\usepackage{orcidlink}
\usepackage{xcolor}
\usepackage{multirow}
\usepackage[normalem]{ulem}

\usepackage[
  protrusion=true,
  expansion=true,
  tracking=true,
  kerning=true,
  spacing=true
]{microtype}

\begin{document}

%\preprint{APS/123-QED}

\title{Universal relation between dipole polarizability of finite nuclei and neutron-star compactness}

\author{P.S. Koliogiannis\orcidlink{0000-0001-9326-7481}$^{1,2}$}
\email{pkoliogi@phy.hr}
\author{T. Ghosh\orcidlink{0000-0002-1794-2817}$^{1}$}
\email{tghosh@phy.hr}
\author{E. Y{\" u}ksel\orcidlink{0000-0002-2892-3208}$^{3}$}
\email{e.yuksel@surrey.ac.uk}
\author{N. Paar\orcidlink{0000-0002-6673-6622}$^{1}$}
\email{npaar@phy.hr}

\affiliation{$^{1}$Department of Physics, Faculty of Science, University of Zagreb, Bijeni\v cka cesta 32, 10000, Zagreb, Croatia}%Lines break automatically or can be forced with \\
\affiliation{$^{2}$Department of Theoretical Physics, Aristotle University of Thessaloniki, 54124 Thessaloniki, Greece}
\affiliation{$^{3}$School of Mathematics and Physics, University of Surrey, Guildford, Surrey, GU2 7XH, United Kingdom}

\date{\today}% It is always \today, today,

%======================================================================================================================
\begin{abstract}
The nuclear equation of state, which determines the structure and properties of neutron stars, remains subject to substantial theoretical uncertainties, leading to model dependence in predicted observables. Universal relations have emerged as a powerful tool to mitigate this dependence by linking neutron star observables in a framework-independent manner. In this work, we introduce a new universal relation that \emph{bridges} finite nuclei and neutron stars through the dimensionless quantity $\zeta = \beta_{1.4}\tilde{L}^{-1}$, which couples the compactness of a $1.4~M_{\odot}$ neutron star to the slope of the nuclear symmetry energy at saturation. The relation is examined under a broad set of relativistic energy density functionals with point-coupling and meson-exchange interactions, as well as non-relativistic Skyrme functionals. We demonstrate that $\zeta$ exhibits a strong exponential correlation with the electric dipole polarizability $\alpha_D$ in finite nuclei across all considered equations of state. By exploiting experimental $\alpha_D$ data for selected neutron-rich nuclei, we constrain $\zeta$ and translate these constraints into equation-of-state independent bounds on the neutron star radius $R_{1.4}$ and the symmetry-energy slope $L$, providing insights into the properties of neutron star matter.
\end{abstract}


\maketitle

%\tableofcontents

%======================================================================================================================
\section{Introduction}
\label{sec:introduction}
%======================================================================================================================
Neutron stars serve as natural laboratories for exploring the interplay between nuclear and gravitational physics under the most extreme conditions in the universe. Their cores reach baryonic densities several times greater than the nuclear saturation density—regimes that remain far beyond the reach of terrestrial experiments. The internal structure and macroscopic properties of neutron stars are governed by the equation of state (EOS) of dense matter, which encapsulates the microphysics through the relationship between pressure and energy density. However, substantial uncertainties in the EOS remain, mainly because the density dependence of the nuclear symmetry energy is poorly known at supra-saturation densities, which are essential for neutron star physics~\cite{Lattimer-2004,Lattimer2021,BURGIO2021103879}.

Over the past decade, multimessenger observations have imposed increasingly stringent constraints on the dense matter EOS. Precision measurements of massive neutron stars with $M \gtrsim 2~M_{\odot}$ establish a robust lower bound on the stiffness of the EOS~\cite{Arzoumanian-2018,Antoniadis-2013,Cromartie-2020,Fonseca_2021,Romani-2022}. Radius estimates from X-ray timing missions, such as NICER, probe the nature of the intermediate density regime~\cite{Miller_2019,Riley_2019,Raaijmakers_2019,Miller_2021,Riley_2021,Dittmann_2024}, while gravitational-wave signals from binary mergers, most notably GW170817 and GW190425, encode tidal deformabilities that directly constrain the EOS at supra-nuclear densities~\cite{Abbott-2019,PhysRevLett.119.161101,Abbott_2020}. Additional constraints arise from rapid rotation which imposes limits on the maximum mass, radius, moment of inertia, quadrupole moment, and dimensionless spin parameter that any viable EOS must satisfy. In this context, the fastest known pulsar, rotating at 716 Hz, serves as an observational benchmark for these limits~\cite{doi:10.1126/science.1123430,10.1111/j.1365-2966.2005.09575.x,PhysRevC.101.015805}.

Complementary constraints on the EOS arise from nuclear physics experiments that probe various properties of finite nuclei~\cite{ROCAMAZA201896}. Measurements of neutron skin thickness in heavy nuclei and dipole polarizabilities provide direct sensitivity to the density dependence of the nuclear symmetry energy around saturation density~\cite{Birkhan2017,PhysRevLett.120.172702,PhysRevLett.111.242503,Rokamaza2015,PhysRevLett.107.062502,PhysRevC.85.041302,Rokamaza2015,PhysRevC.92.031305,Bassauer2020,PhysRevC.111.024312}, as well as at subsaturation densities \cite{PhysRevC.92.031301}. In parallel, heavy-ion collision observables offer insights into the pressure of asymmetric matter at supra-saturation densities~\cite{PhysRevC.94.034608,LEFEVRE2016112,Huth_2022}. These measurements, in conjunction with astrophysical constraints, aim to confine the EOS across densities and establish a firm link between finite nuclei and neutron star properties.\\
\indent In addition to observational and experimental bounds, an important development has been the discovery of universal relations among bulk neutron star properties. Being largely independent of the underlying EOS, these relations enable connections between otherwise inaccessible stellar observables, thereby constrain the behavior of dense matter at supra-nuclear densities. Among these, the I-Love-Q set represents the foundational relation, connecting the moment of inertia, tidal deformability, and quadrupole moment~\cite{doi:10.1126/science.1236462,PhysRevD.88.023007,PhysRevD.88.023009,PhysRevLett.112.201102,PhysRevD.101.124006,PhysRevD.111.024075}. The neutron star binding energy has been shown to obey universal relations with compactness and tidal deformability, providing an additional EOS-insensitive observable for both isolated and binary systems~\cite{Lattimer_2001,PhysRevD.102.103011,universe8080395,LATTIMER2007109}. These correlations have been further extended to a wide range of scenarios, including rapidly rotating stars~\cite{Doneva_2014,PhysRevD.90.064026,Breu_2016,PhysRevD.96.024046,Luk_2018,PhysRevD.99.043004,PhysRevC.101.015805,PhysRevD.102.023039,PhysRevD.103.063038,Koliogiannis_2021,PhysRevD.107.103050,Musolino_2024,PhysRevD.109.103033}, finite-temperature stars relevant for mergers~\cite{PhysRevD.90.064026,PhysRevC.96.045806,Raduta_2020,PhysRevC.103.055811,universe8080395}, and hybrid configurations that incorporate additional degrees of freedom such as deconfined quarks or hyperons~\cite{Raduta_2020,doi:10.1126/science.1236462,YAGI20171,Wei_2019,Khosravi_2022,Kumar_2024}, as well as stability conditions~\cite{Koliogiannis_2019}. Universal relations have also been applied to gravitational-wave asteroseismology, allowing oscillation mode frequencies to constrain neutron star masses and radii~\cite{PhysRevLett.77.4134,10.1046/j.1365-8711.1998.01840.x,PhysRevD.88.044052,PhysRevD.104.043011}. Furthermore, combining astrophysical observations with nuclear physics–informed quantities has been used to tighten constraints on neutron star bulk properties~\cite{PhysRevD.88.023009,Yagi_2017,PhysRevD.99.123026,PhysRevD.106.043027,10.1093/mnras/stad1967,PhysRevD.107.023010,PhysRevD.108.104065,PhysRevD.110.024011,PhysRevD.109.103029,PhysRevD.109.023020,Chatterjee_2025}. In binary neutron star systems, universal relations for tidal deformabilities enable gravitational-wave measurements of the inspiral phase to more directly constrain neutron star structure and the EOS via tidal effects encoded in the waveform~\cite{PhysRevD.104.023005,PhysRevD.103.063036,PhysRevD.107.043010}.\\
\indent In this work, we introduce a universal relation connecting a finite-nucleus observable—the electric dipole polarizability, $\alpha_D$—to a key neutron star property, namely the compactness of a canonical $1.4~M_{\odot}$ neutron star. Both quantities are strongly influenced by the slope of the nuclear symmetry energy at saturation, $L$, which governs the pressure of neutron-rich matter~\cite{ROCAMAZA201896}, thereby establishing a direct connection between nuclear experiments and neutron star structure. To explore this relation, we employ a comprehensive set of microscopic nuclear models based on energy density functional (EDF) theory, including relativistic density-dependent point-coupling (DD--PC)~\cite{Yuksel2021,KOLIOGIANNIS2025139362}, density-dependent meson-exchange (DD--ME)~\cite{PhysRevC.68.024310}, and nonlinear meson-exchange (NL)~\cite{Reinhard1986,PhysRevC.55.540} interactions, alongside non-relativistic Skyrme functionals~\cite{PhysRevC.109.055801}. EDFs provide a unified and self-consistent framework for describing both finite nuclei and neutron stars, capturing bulk nuclear properties and the density dependence of the symmetry energy that are essential for establishing robust nuclear–astrophysical correlations. The resulting universal relation exhibits a remarkably consistent exponential trend across the full set of EOSs and nuclei considered, allowing us to place constraints on neutron star radii and the symmetry energy slope. Furthermore, it serves as a predictive tool for $\alpha_D$ in nuclei where experimental measurements are not yet available. Altogether, this framework provides a direct, largely model-independent bridge between terrestrial nuclear experiments and astrophysical observations, offering a stringent benchmark for the nuclear EOS.

The paper is organized as follows. Section~\ref{sec:nuc_models} presents the nuclear models and the corresponding EOSs employed in this work, while Sec.~\ref{sec:methodology} lays out the followed methodology. Section~\ref{sec:results} contains the results and their discussion, where we focus on the universal relation, and implications for neutron star and finite nuclei structure. Finally, Sec.~\ref{sec:conclusions} summarizes the main findings and concludes the implications for nuclear and astrophysical observables.

%======================================================================================================================
\section{Nuclear Models and Equations of State}
\label{sec:nuc_models}
%======================================================================================================================
In this work, we employ a comprehensive set of relativistic and non-relativistic EDFs, which are capable of consistently describing finite nuclei and neutron-star matter. This framework enables a systematic exploration of universal relations linking microscopic nuclear structure to macroscopic stellar properties. Within the relativistic EDFs, we consider several variants of DD--PC~\cite{Yuksel2021,KOLIOGIANNIS2025139362}, DD--ME~\cite{PhysRevC.68.024310,PhysRevC.71.024312}, and NL~\cite{Reinhard1986,PhysRevC.55.540} interactions. The non-relativistic EDFs include Skyrme parameterizations—KDEv01~\cite{PhysRevC.72.014310}, SGI~\cite{VANGIAI1981379}, SK255 and SK272~\cite{PhysRevC.68.031304}, SLy230a, SLy4, and SLy5~\cite{CHABANAT1997710,CHABANAT1998231}, as well as SkI2, SkI3, and SkI5~\cite{REINHARD1995467}. In addition, we employ the SAMi-J family of Skyrme functionals~\cite{PhysRevC.86.031306} (SAMi-J30, SAMi-J32, and SAMi-J34), which is specifically designed to allow a controlled variation of the symmetry-energy slope.

\begin{table}
	%\squeezetable
	\caption{Properties of the EOSs, including the incompressibility $K$ (MeV), the symmetry energy $J$ (MeV), the slope of the symmetry energy $L$ (MeV), the maximum mass $M_{\rm max}$ $(M_{\odot})$, and the corresponding radius $R_{\rm max}$ (km), and the radius of a $1.4~M_{\odot}$ neutron star $R_{1.4}$ (km).}
	\begin{ruledtabular}
        \begin{tabular}{lrrrrrr}
            EOS & $K$ & $J$ & $L$ & $M_{\rm max}$ & $R_{\rm max}$ & $R_{1.4}$ \\
            \hline
            DD--PC-J29  & 230.00 & 29.00 & 29.00 & 2.20 & 10.61 & 11.81 \\
            DD--PC-J30  & 230.00 & 30.00 & 35.60 & 2.20 & 10.63 & 11.92 \\
            DD--PC-J31  & 230.00 & 31.00 & 43.80 & 2.19 & 10.63 & 12.00 \\
            DD--PC-J32  & 230.00 & 32.00 & 52.30 & 2.19 & 10.63 & 12.08 \\
            DD--PC-J33  & 230.00 & 33.00 & 62.00 & 2.18 & 10.63 & 12.21  \\
            DD--PC-J34  & 230.00 & 34.00 & 72.10 & 2.17 & 10.65 & 12.38 \\
            DD--PC-J35  & 230.00 & 35.00 & 83.20 & 2.17 & 10.71 & 12.62  \\
            DD--PC-J36  & 230.00 & 36.00 & 94.10 & 2.17 & 10.83 & 12.90  \\
            DD--PC-CREX  & 225.48 & 27.01 & 19.60 & 2.09 & 10.06 & 11.26 \\
            DD--PC-PREX  & 235.41 & 36.18 & 101.78 & 2.07 & 10.44 & 12.75 \\
            DD--PC-REX  & 242.95 & 28.86 & 30.03 & 2.19 & 10.56 & 11.82  \\
            DD--PCX  & 213.03 & 31.12 & 46.32 & 2.16 & 10.45 & 11.86 \\
            DD--ME-J30  & 250.00 & 30.00 & 30.01 & 2.47 & 11.96 & 12.87 \\
            DD--ME-J32  & 250.00 & 32.00 & 46.53 & 2.46 & 11.98 & 13.11 \\
            DD--ME-J34  & 250.00 & 34.00 & 62.06 & 2.44 & 11.99 & 13.36  \\
            DD--ME-J36  & 250.00 & 36.00 & 85.43 & 2.44 & 12.12 & 13.82 \\
            DD--ME-J38  & 250.00 & 38.00 & 110.64 & 2.48 & 12.43 & 14.39 \\
            DD--ME1  & 244.64 & 34.02 & 55.43 & 2.44 & 11.94 & 13.22 \\
            DD--ME2  & 250.83 & 33.20 & 51.26 & 2.48 & 12.07 & 13.25  \\
            NL-SC  & 222.98 & 35.22 & 108.31 & 2.64 & 12.75 & 14.44 \\
            NL-SH  & 355.26 & 36.11 & 113.51 & 2.80 & 13.54 & 14.98 \\
            NL1  & 211.02 & 43.46 & 139.93 & 2.81 & 13.40 & 14.92 \\
            NL3  & 271.65 & 37.41 & 118.45 & 2.77 & 13.34 & 14.85 \\
            NL3$^{*}$  & 258.18 & 38.68 & 122.50 & 2.76 & 13.26 & 14.80 \\
            NLC  & 223.00 & 34.89 & 107.32 & 2.64 & 12.75 & 14.43 \\
            KDEv01  & 227.53 & 34.58 & 54.70 & 1.97 & 9.77 & 11.58 \\
            SAMi-J30  & 245.00 & 30.00 & 63.18 & 2.14 & 10.60 & 12.55 \\
            SAMi-J32  & 245.00 & 32.00 & 85.10 & 2.18 & 11.01 & 13.27 \\
            SAMi-J34  & 245.00 & 34.00 & 105.31 & 2.19 & 11.25 & 13.83 \\
            SGI  & 261.74 & 28.33 & 63.86 & 2.24 & 10.96 & 12.87 \\
            SK255  & 254.92 & 37.40 & 95.05 & 2.14 & 10.86 & 13.23 \\
            SK272  & 271.50 & 37.40 & 91.67 & 2.23 & 11.08 & 13.36 \\
            SLy230a  & 229.87 & 31.99 & 44.32 & 2.11 & 10.30 & 11.90 \\
            SLy4  & 229.90 & 32.00 & 45.96 & 2.05 & 9.99 & 11.69 \\
            SLy5  & 229.92 & 32.03 & 48.27 & 2.05 & 10.04 & 11.76 \\
            SkI2  & 240.92 & 33.37 & 104.33 & 2.16 & 11.17 & 13.70 \\
            SkI3  & 258.18 & 34.83 & 100.52 & 2.24 & 11.35 & 13.69 \\
            SkI5  & 255.78 & 36.64 & 129.33 & 2.24 & 11.54 & 14.36 \\
        \end{tabular}
    \end{ruledtabular}
	\label{tab:table1}
\end{table}

Overall, this study employs EOSs derived from 12 DD--PC, 7 DD--ME, 6 NL, and 13 Skyrme functionals, covering a broad range of nuclear matter properties, including incompressibility 
$K \sim 211$–$356$ MeV, symmetry energy 
$J \sim 27$–$44$ MeV, and its slope 
$L \sim 19$–$140$ MeV. 
Neutron star matter is described across three density regimes: 
(i) the liquid core, 
(ii) the inner crust, and (iii) the outer crust. 
The liquid core consists of neutrons, protons, electrons, and muons in $\beta$-equilibrium and charge neutrality. 
For densities relevant to the crust, we adopt the SLy EOS~\cite{Douchin_2001} for the inner crust and the BPS~\cite{Baym-71} and FMT~\cite{PhysRev.75.1561} EOSs for the outer crust, matched to the core EOS at the crust--core transition determined using the thermodynamical method~\cite{PhysRevC.85.024302,PhysRevC.90.011304}. 

\begin{figure*}
    \centering
    \includegraphics[width=\textwidth]{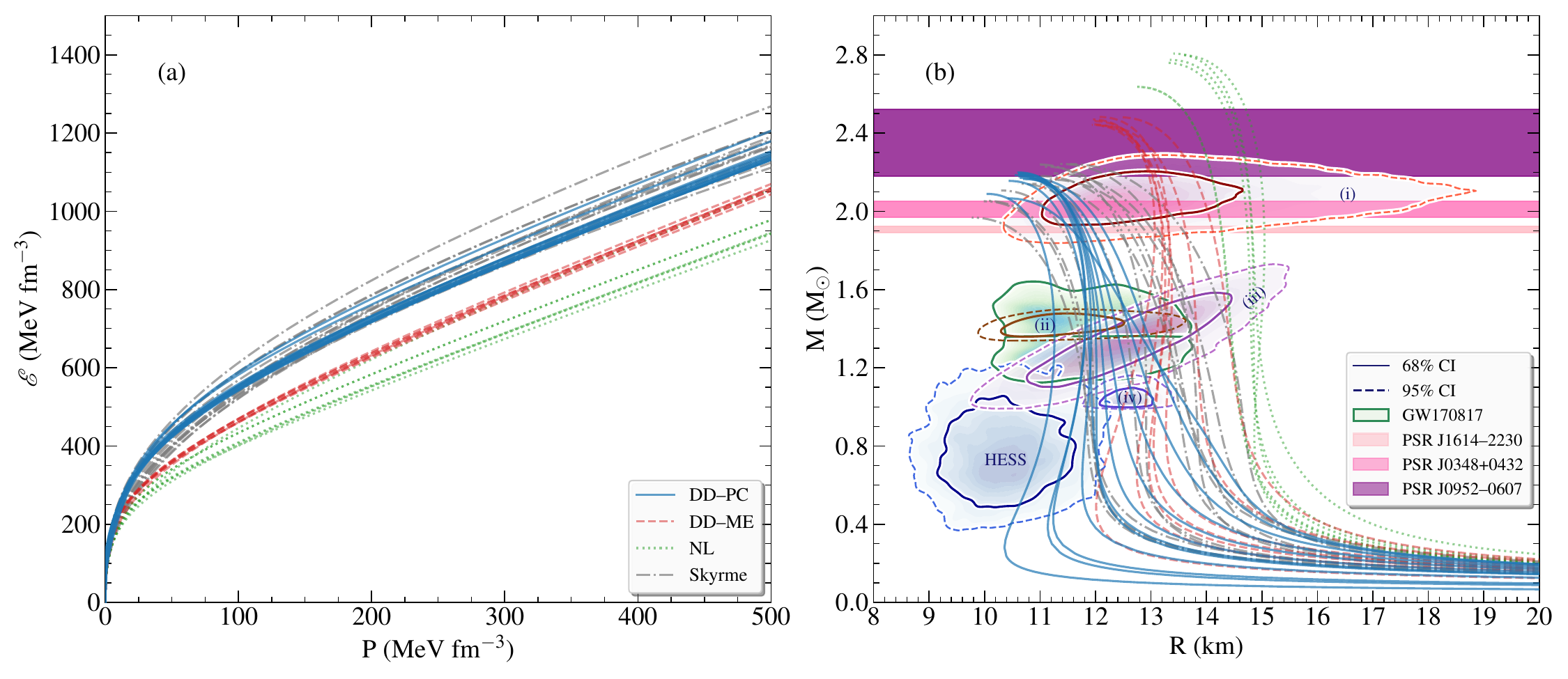}
    \caption{(a) The energy density as a function of pressure for the full set of EOSs. (b) The corresponding gravitational mass as a function of radius for the full set of EOSs. The shaded contour labeled HESS indicates constraints inferred from the HESS J1731--347 supernova remnant~\cite{Doroshenko-2022}. Additional observational constraints are shown for: (i) PSR J0740+6620~\cite{Fonseca_2021,Dittmann_2024}, (ii) PSR J0437--4715~\cite{Choudhury_2024}, (iii) PSR J0030+0451~\cite{Riley_2019,Raaijmakers_2019}, and (iv) PSR J1231--1411~\cite{Salmi_2024}. Horizontal shaded bands denote mass measurements of the massive pulsars PSR J1614--2230~\cite{Arzoumanian-2018}, PSR J0348+0432~\cite{Antoniadis-2013}, and PSR J0952--0607~\cite{Romani-2022}, providing lower bounds on the maximum neutron star mass. Constraints from the binary neutron star merger GW170817~\cite{Abbott-2019} are also indicated.}
    \label{fig:MR}
\end{figure*}

Table~\ref{tab:table1} presents the full set of EOSs used in this work, along with key nuclear matter and neutron star properties, including the incompressibility $K$, the symmetry energy $J$, the slope of the symmetry energy $L$, the maximum mass $M_{\rm max}$, the corresponding radius $R_{\rm max}$, and the radius at $1.4~M_{\odot}$, $R_{1.4}$. All EOSs are calibrated to reproduce the saturation properties of symmetric nuclear matter and support a maximum neutron star mass of at least $2~M_{\odot}$, in agreement with current astrophysical observations. 

The wide diversity of nuclear matter properties, together with the substantial spread in the predicted radii for $1.4~M_{\odot}$ neutron stars (approximately 11–15 km), enables a comprehensive and systematic exploration from soft toward stiff models. This broad coverage ensures that the correlations under investigation, linking nuclear observables to neutron-star properties, remain independent of the underlying stiffness of the EOS. Furthermore, by encompassing both extremely soft and stiff scenarios, the analysis provides a stringent probe of these correlations, demonstrating their robustness across the full spectrum of neutron-stars~\cite{doi:10.1126/science.1236462,PhysRevD.88.023007,PhysRevD.88.023009,PhysRevLett.112.201102,PhysRevD.101.124006,PhysRevD.111.024075,Lattimer_2001,PhysRevD.102.103011,universe8080395,LATTIMER2007109,Doneva_2014,PhysRevD.90.064026,Breu_2016,PhysRevD.96.024046,Luk_2018,PhysRevD.99.043004,PhysRevC.101.015805,PhysRevD.102.023039,PhysRevD.103.063038,PhysRevD.107.103050,Musolino_2024,Koliogiannis_2021,PhysRevD.109.103033,PhysRevC.96.045806,Raduta_2020,PhysRevC.103.055811,YAGI20171,Wei_2019,Khosravi_2022,Kumar_2024,Koliogiannis_2019,PhysRevLett.77.4134,10.1046/j.1365-8711.1998.01840.x,PhysRevD.88.044052,PhysRevD.104.043011,Yagi_2017,PhysRevD.99.123026,PhysRevD.106.043027,10.1093/mnras/stad1967,PhysRevD.107.023010,PhysRevD.108.104065,PhysRevD.110.024011,PhysRevD.109.103029,PhysRevD.109.023020,Chatterjee_2025,PhysRevD.104.023005,PhysRevD.103.063036,PhysRevD.107.043010}. For illustration, the associated energy–pressure and mass–radius relations are shown in Fig.~\ref{fig:MR}, overlaid with up-to-date astrophysical constraints~\cite{Arzoumanian-2018,Antoniadis-2013,Cromartie-2020,Fonseca_2021,Romani-2022,Miller_2019,Riley_2019,Raaijmakers_2019,Miller_2021,Riley_2021,Dittmann_2024,Abbott-2019,Doroshenko-2022,Salmi_2024,Choudhury_2024,Abbott-2019}.

%======================================================================================================================
\section{Methodology}
\label{sec:methodology}
%======================================================================================================================
To establish a quantitative link between nuclear structure and neutron star properties, we focus on two key observables:  
(a) the electric dipole polarizability, $\alpha_D$, a sensitive isovector observable characterizing the nuclear response to an external electric field~\cite{ROCAMAZA201896,Koliogiannis_2026}, and  
(b) the stellar compactness, $\beta$, which encodes the interplay between gravitational mass and radius of neutron stars.

The electric dipole polarizability $\alpha_D$ is defined as the inverse energy-weighted sum of the isovector dipole response in finite nuclei, largely governed by the isovector giant dipole resonance, a collective oscillation of neutrons against protons~\cite{Paar_2007,Rocamaza2013,ROCAMAZA201896}. It is calculated as
\begin{equation}
    \label{eq:alphaD}
    \alpha_{D}=\frac{8\pi e^{2}}{9}\int_{0}^{\infty}E^{-1}S(E1;E)dE,
\end{equation}
where $S(E1;E)$ denotes the electric dipole strength as a function of the excitation energy $E$, obtained using the quasiparticle random phase approximation in both relativistic~\cite{PhysRevC.67.034312,Paar_2007} and non-relativistic frameworks~\cite{colo2013self,colo2021userguidehfbcsqrpav1code}. On the other hand, stellar compactness $\beta$ is a dimensionless quantity characterizing the neutron star and is defined as
\begin{equation}
    \beta = \frac{GM}{c^2R},
    \label{eq:compactness}
\end{equation}
where $M$ and $R$ denote the gravitational mass and the radius, respectively.
Both dipole polarizability $\alpha_D$ and stellar compactness $\beta$ are strongly governed by the density dependence of the symmetry energy slope, $L$, which acts as a natural mediator connecting nuclear and stellar properties: $\alpha_D$ correlates directly with $L$ (see also Refs.~\cite{PhysRevC.81.051303, Koliogiannis_2026}), while $\beta$ reflects its influence via the neutron star radius. In particular, the well-established correlation between $R_{1.4}$ and the pressure of neutron star matter up to $\sim$$2\rho_{0}$~\cite{Lattimer_2001}, extended to $R_{1.4}$ and $L$~\cite{PhysRevC.109.055801}, ensures that $L$ provides a link across the relevant density regimes.

\begin{figure*}
    \centering
    \includegraphics[width=\textwidth]{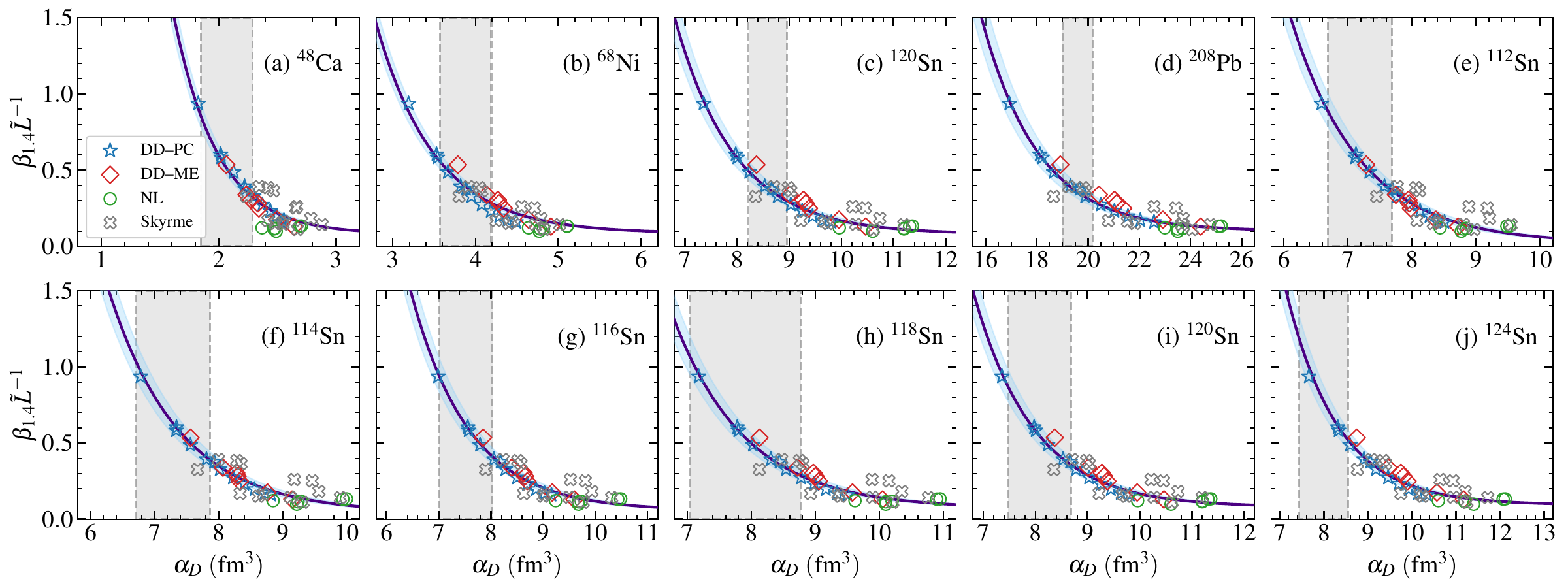}
    \caption{The dimensionless quantity $\zeta \equiv \beta_{1.4}\,\tilde{L}^{-1}$ as a function of the electric dipole polarizability $\alpha_D$ for the full set of EOSs. Vertical shaded regions indicate experimental values of $\alpha_D$ for: (a) $^{48}$Ca~\cite{Birkhan2017}, (b) $^{68}$Ni~\cite{PhysRevLett.111.242503}, (c) $^{120}$Sn~\cite{Rokamaza2015,PhysRevC.92.031305}, (d) $^{208}$Pb~\cite{Rokamaza2015,PhysRevLett.107.062502}, and (e--j) $^{112-124}$Sn~\cite{Bassauer2020}. The solid curve represents the universal relation~\eqref{eq:un_rel}, with the shaded band indicating its associated uncertainty.}
    \label{fig:a_D_slope}
\end{figure*}

To formalize this connection, we introduce a dimensionless quantity
\begin{equation}
\zeta \equiv \beta_{1.4}\,\tilde{L}^{-1},
\end{equation}
where $\beta_{1.4}$ denotes the compactness of a $1.4~M_{\odot}$ neutron star, and $\tilde{L}=L/L_0$ with $L_0 = 100$ MeV as a reference value. The ratio $\zeta$ effectively characterizes the stiffness of the EOS: lower values correspond to stiffer models with greater pressure support, where the $L$ and $R_{1.4}$ values are higher, whereas higher values indicate softer behavior. Specifically, $L$ probes the EOS near saturation density $\rho_0$, while $\beta_{1.4}$ captures the influence of the symmetry energy slope at densities $\sim$$(2-3)\rho_0$ through its dependence on $R_{1.4}$, which can be related to saturation density using experimental data from finite nuclei~\cite{PhysRevLett.120.172702}. Since $\beta_{1.4}$ and $L$ probe complementary density regimes, their combination provides a physically motivated bridge between neutron star structure and finite nuclei, establishing $\zeta$ as a natural candidate for exploring universal correlations with the electric dipole polarizability, $\alpha_D$.

%======================================================================================================================
\section{Results and discussion}
\label{sec:results}
%======================================================================================================================
%======================================================================================================================
\subsection{Universal finite nuclei -- Neutron star relation}
\label{sec:universal_relation}
%======================================================================================================================
Figure~\ref{fig:a_D_slope} illustrates the dimensionless quantity $\zeta$ as a function of the electric dipole polarizability $\alpha_D$ for the full set of EOSs, as well as for ten neutron-rich nuclei for which experimental $\alpha_D$ values are available:
$^{48}$Ca~\cite{Birkhan2017},  
$^{68}$Ni~\cite{PhysRevLett.111.242503},  
$^{120}$Sn~\cite{Rokamaza2015,PhysRevC.92.031305},  
$^{208}$Pb~\cite{Rokamaza2015,PhysRevLett.107.062502},  
and even-even isotopes $^{112-124}$Sn~\cite{Bassauer2020}. Despite their different theoretical foundations, all EOS-models follow a general trend, revealing an EOS-insensitive correlation between dipole polarizability and neutron-star compactness, driven primarily by the symmetry energy slope. We find that this trend is well reproduced by the following empirical exponential relation, chosen as a simple functional form that accurately captures the observed nonlinear behavior over the full range of $\alpha_D$ values considered:
\begin{equation}
    \zeta(\alpha_D,A,\delta)
    = c_{1}(A)e^{-c_{2}(A)\alpha_D} + c_{3}(\delta),
    \label{eq:un_rel}
\end{equation}
where the coefficients $c_{i}$ $(i=1-3)$ depend on the nuclear mass number $A$ and the isospin asymmetry $\delta=(N-Z)/A$, with $N$ and $Z$ denoting the neutron and proton numbers, respectively. These coefficients encode mass and isospin systematics relevant to both finite nuclei and neutron-star properties.
Here, we introduce the following expansions of the coefficients,
\begin{eqnarray}
    c_{1}(A) &=& \sum_{k=0}^{3} q_{1k}A^{-k}, \label{eq:c1} \\
    c_{2}(A) &=& \sum_{k=0}^{3} q_{2k}A^{-k}, \label{eq:c2} \\
    c_{3}(\delta) &=& \sum_{k=0}^{3} q_{3k}\delta^{k}. \label{eq:c3}
\end{eqnarray}
The expansions in Eqs.~(\ref{eq:c1})--(\ref{eq:c3}) are truncated at \(k=3\), a choice that balances accuracy with simplicity by providing an excellent description of the nuclear and neutron star data without introducing unnecessary fit parameters. The inverse-power dependence on $A$ captures the smooth mass variation of bulk nuclear properties, while the polynomial dependence on $\delta$ accounts for residual isospin effects. Truncation at higher-order terms, $k>3$, do not yield appreciable improvement, ensuring a stable and robust parametrization.\\
\indent Although Fig.~\ref{fig:a_D_slope} displays results only for nuclei with measured $\alpha_D$ values to avoid redundancy, the fit, given by Eq.~\eqref{eq:un_rel}, was performed using a broader training set of fourteen nuclei:  
$^{48}$Ca, $^{54}$Ca, $^{60}$Ca, $^{86}$Kr, $^{68}$Ni, $^{78}$Ni, $^{208}$Pb, $^{112\text{--}124}$Sn, and $^{96}$Zr, spanning a wide range in nuclear mass number ($A\sim 48$--$208$) and isospin asymmetry ($\delta\sim 0.11$--$0.33$). The resulting regression achieves a global coefficient of determination $R^2 \gtrsim 0.9$, indicating that the relation reliably describes both light and heavy nuclei. The extracted expansion coefficients $q_{ik}$, with $i=1$--$3$, are listed in Table~\ref{tab:table2}.

\begin{table}
	\caption{Coefficients of the expansions in Eqs.~\eqref{eq:c1}-\eqref{eq:c3} that appear in Eq.~\eqref{eq:un_rel}.}
	\begin{ruledtabular}
        \begin{tabular}{lrrr}
            Coefficients & $q_{1k}$ & $q_{2k}~{\rm (fm^{-3})}$ & $q_{3k}$ \\
            \hline
            \multicolumn{4}{c}{\vspace{-0.25cm}} \\
            $k=0$ &  5.558$\times 10^{3}$ & -0.770 & -0.257 \\
            $k=1$ &   -1.080$\times 10^{6}$ & 3.666$\times 10^{2}$ & 3.993 \\
            $k=2$ &  6.909$\times 10^{7}$ & -2.749$\times 10^{4}$ & -15.288 \\
            $k=3$ &  -1.432$\times 10^{9}$ & 8.540$\times 10^{5}$ & 20.714 \\
        \end{tabular}
    \end{ruledtabular}
	\label{tab:table2}
\end{table}

%======================================================================================================================
\subsection{Neutron star constraints}
\label{sec:ns_constraints}
%======================================================================================================================
The existence of the universal relation given in Eq.~\eqref{eq:un_rel}, alongside with the precise experimental measurements of the electric dipole polarizability $\alpha_D$ in ten neutron-rich nuclei~\cite{Birkhan2017,PhysRevLett.111.242503,Rokamaza2015,PhysRevC.92.031305,PhysRevLett.107.062502,Bassauer2020}, indicated by the vertical shaded bands in Fig.~\ref{fig:a_D_slope}, allows us to impose quantitative bounds on the dimensionless quantity $\zeta$. To this end, we follow the procedure introduced in Ref.~\cite{Koliogiannis_2026}, in which the nuclei are divided into two subsets in order to account for differences in the completeness and reliability of the experimental dipole response.

Specifically, the first subset, denoted CNSP-4, consists of $^{48}$Ca, $^{68}$Ni, $^{120}$Sn, and $^{208}$Pb, for which $\alpha_D$ has been extracted with minimal theoretical input. The second subset, CNSP-10, extends this selection by including the even-even Sn isotopes $^{112\text{--}124}$Sn, thereby providing a broader experimental basis. For these isotopes, however, contributions to $\alpha_D$ above 20~MeV are reconstructed using quasiparticle--phonon model calculations~\cite{Bassauer2020}, introducing an additional level of model dependence. Recent analyses indicate that these reconstructed strengths tend to favor lower values of the symmetry energy compared to those inferred from other nuclei, resulting in a mild tension with existing constraints. For this reason, CNSP-4 is regarded as the more conservative and reliable set for neutron star applications, although results obtained using CNSP-10 are also reported for completeness (see Ref.~\cite{Koliogiannis_2026} for a detailed discussion).

Using the nuclei in each subset, we calculate weighted averages of $\zeta$, obtaining
\begin{align}
    \text{CNSP-4}: \quad \zeta &= 0.390 \pm 0.052, \\
    \text{CNSP-10}: \quad \zeta &= 0.435 \pm 0.047 .
\end{align}
These bounds delineate the region permitted by the interplay between the compactness of a $1.4~M_{\odot}$ neutron star and the slope of the symmetry energy. The higher $\zeta$ value obtained for CNSP-10 reflects its tendency toward smaller symmetry-energy slopes, consistent with the discussion above.

While $\zeta$ encapsulates the interplay between nuclear and neutron-star physics in a compact and model-independent form, its physical content can be further explored through its constituents, the neutron-star radius $R_{1.4}$ and the slope of the symmetry energy $L$. Guided by the bounds on $\zeta$, we therefore derive corresponding constraints on the product $R_{1.4}L$. In particular, we find
\begin{align}
\text{CNSP-4}: \quad R_{1.4}L &= 554.4 \pm 73.4\,({\rm km\,MeV}), \label{eq:RL_limit_cnsp4} \\
\text{CNSP-10}: \quad R_{1.4}L &= 440.6 \pm 47.9\,({\rm km\,MeV}), \label{eq:RL_limit_cnsp10}
\end{align}
which define two experimentally informed bands that overlap within a relatively narrow region. Although these products do not uniquely determine $R_{1.4}$ without independent information on $L$, they nevertheless impose nontrivial constraints that any viable EOS must satisfy. In this sense, the quantity $R_{1.4}L$ emerges as a nuclear-physics–informed consistency criterion for neutron star models.

\begin{figure}[t!]
    \centering
    \includegraphics[width=\columnwidth]{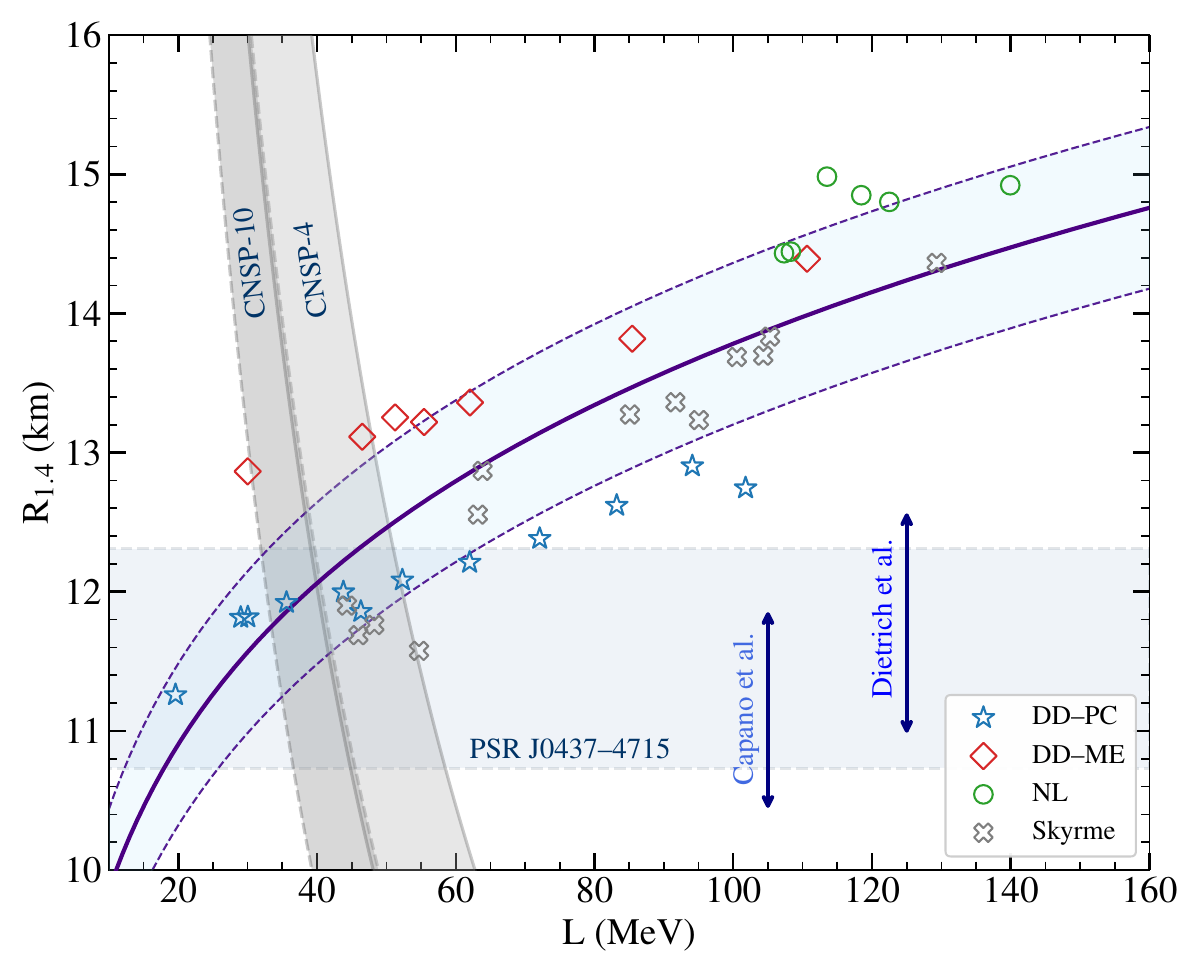}
    \caption{The radius of a canonical $1.4~M_{\odot}$ neutron star, $R_{1.4}$, as a function of the slope of the symmetry energy $L$ for the full set of EOSs. Curved shaded regions indicate constraints inferred from the CNSP-4 and CNSP-10 sets of nuclei. The solid curve shows the systematic trend described by Eq.~\eqref{eq:radius_L}, with the associated uncertainty represented by the shaded band. Vertical arrows denote neutron-star radius constraints from Refs.~\cite{Capano-2020,doi:10.1126/science.abb4317}, while the horizontal shaded region indicates the radius of PSR J0437--4715~\cite{Choudhury_2024}.}
    \label{fig:RL}
\end{figure}

Figure~\ref{fig:RL} displays $R_{1.4}$ as a function of the symmetry energy slope $L$ for the full set of EOSs. As mentioned in Sec.~\ref{sec:methodology}, we also denote the high-degree correlation of $R_{1.4}$ and $L$, expressed as 
\begin{equation}
    R_{1.4} = 7.049(L/{\rm MeV})^{0.146} \pm 0.581~{\rm (km)}.
    \label{eq:radius_L}
\end{equation}
In addition, we overlay the $\alpha_D-$informed constraints, CNSP-4 and CNSP-10 from Eqs.~\eqref{eq:RL_limit_cnsp4} and~\eqref{eq:RL_limit_cnsp10}, respectively. As these bands serve as a consistency criterion for neutron stars, their intersection with Eq.~\eqref{eq:radius_L} defines the corresponding bounds on both $R_{1.4}$ and $L$, shown in Table~\ref{tab:table4}. The results indicate that CNSP-4 and CNSP-10 constraint bands confine the allowed region in $R_{1.4}$--$L$ plane to a relatively narrow domain compared to the broad distribution spanned by the full set of EOSs. This underscores the ability of the proposed framework to significantly reduce the viable parameter space, effectively isolating a subset of EOSs consistent with dipole polarizability constraints. Notably, this constrained region remains compatible with current observational bounds on neutron-star radii, further reinforcing the robustness and physical relevance of the correlation~\cite{Choudhury_2024,Capano-2020,doi:10.1126/science.abb4317}.

\begin{figure*}
    \centering
    \includegraphics[width=\textwidth]{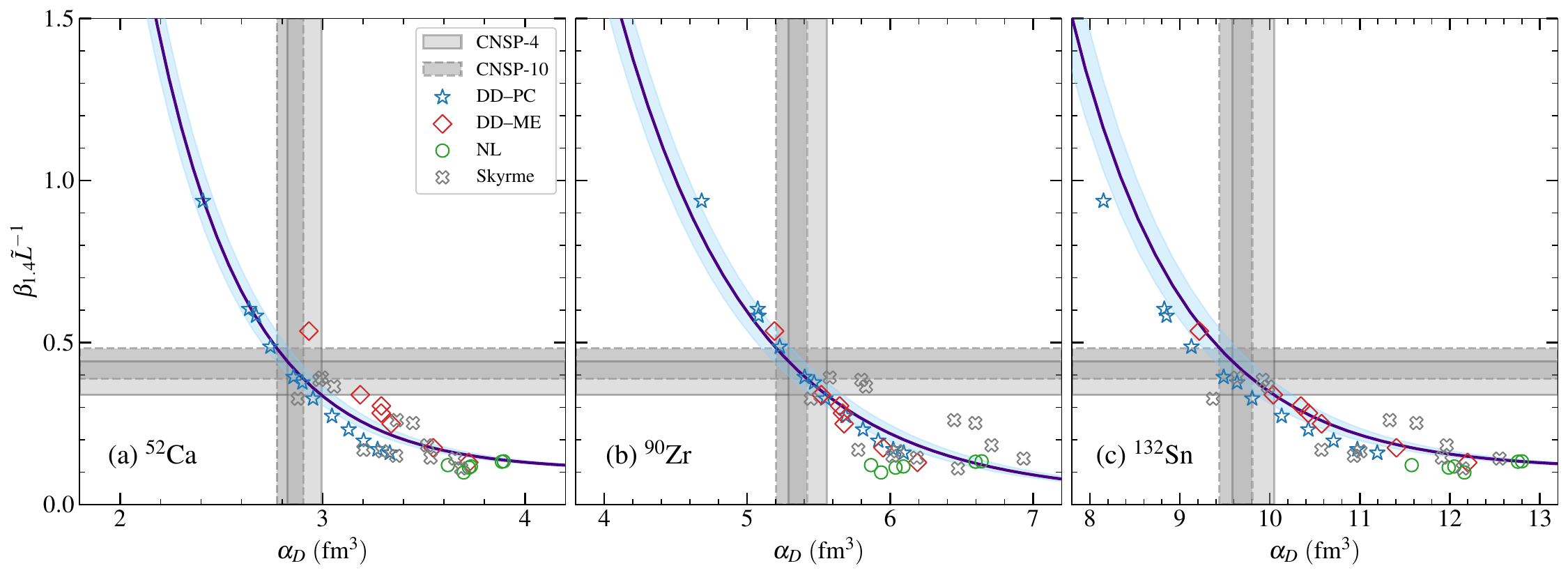}
    \caption{The dimensionless quantity $\zeta \equiv \beta_{1.4}\tilde{L}$ as a function of the electric dipole polarizability $\alpha_D$ for the full set of EOSs, shown for (a) $^{52}$Ca, (b) $^{90}$Zr, and (c) $^{132}$Sn. The solid curve represents the universal relation~\eqref{eq:un_rel}, with the associated uncertainty indicated by the shaded band. Horizontal shaded regions indicate constraints inferred from the CNSP-4 and CNSP-10 sets of nuclei, while vertical shaded regions indicate the corresponding bounds on the dipole polarizability.}
    \label{fig:pred}
\end{figure*}

\begin{table}
    \caption{Constraints on neutron star properties derived from the CNSP-4 and CNSP-10 sets of nuclei. 
    The values correspond to the ranges obtained for the radius of a $1.4~M_{\odot}$ neutron star ($R_{1.4}$) and the slope of the symmetry energy ($L$).}
    \footnotesize
    \begin{ruledtabular}
    \begin{tabular}{lccccc}
        \multicolumn{1}{c}{Nuclei set} & 
        \multicolumn{2}{c}{$R_{1.4}$ (km)} & 
        \multicolumn{2}{c}{$L$ (MeV)} \\
        & min & max & min & max \\
        \hline
        \multicolumn{5}{c}{\vspace{-0.25cm}} \\
        CNSP-4 & $12.057^{+0.505}_{-0.508}$ & $12.472^{+0.505}_{-0.508}$ & $39.896^{+1.615}_{-1.746}$ & $50.335^{+1.972}_{-2.126}$ \\
        \multicolumn{5}{c}{\vspace{-0.25cm}} \\
        CNSP-10 & $11.750^{+0.505}_{-0.509}$ & $12.080^{+0.505}_{-0.508}$ & $33.418^{+1.386}_{-1.502}$ & $40.437^{+1.633}_{-1.766}$
    \end{tabular}
    \end{ruledtabular}
    \label{tab:table4}
\end{table}

The two constraint bands reflect not only the experimental uncertainties but also the differing physical implications associated with each nuclear set. The CNSP-4 set, which relies exclusively on measured dipole strength distributions, favors moderately larger radii and symmetry energy slopes. In contrast, the CNSP-10 set, influenced by the model-dependent treatment of high-energy dipole strength in the Sn isotopes, leads to systematically smaller values of $R_{1.4}$ and $L$. This distinction highlights the sensitivity of neutron star constraints to the details of the nuclear dipole response and indicates that future high-precision measurements of $\alpha_D$, particularly in medium and heavy nuclei, will play an important role in further reducing the associated uncertainties.

%======================================================================================================================
\subsection{Predictions of \texorpdfstring{$\alpha_D$}{alphaD} for a set of finite nuclei}
\label{sec:aD_predictions}
%======================================================================================================================
An important consequence of first establishing the universal relation in Eq.~\eqref{eq:un_rel} and then constraining the dimensionless quantity $\zeta$ is its predictive power. Once experimental data bound $\zeta$, the relation can be inverted to predict the electric dipole polarizability $\alpha_D$ of finite nuclei for which direct measurements are not yet available.

In this work, we focus on three nuclei for which experimental measurements of the electric dipole polarizability are not yet available: $^{52}$Ca, $^{90}$Zr, and $^{132}$Sn. These nuclei were not included in the construction of Eq.~\eqref{eq:un_rel} and span a wide range of nuclear masses and isospin asymmetries. They probe different shell closures and exhibit distinct pairing characteristics, providing a stringent test of the predictive power of the proposed universal relation in regions of the nuclear chart not constrained by experimental input. In particular,
\begin{itemize}
    \item[(a)] \text{$^{52}$Ca}: a neutron-rich nucleus with a subshell closure at $N=32$, exhibiting enhanced shell effects and sensitivity to the symmetry energy in the low-mass region. Its emerging magicity makes it an excellent test case for isovector properties.
    \item[(b)] \text{$^{90}$Zr}: an intermediate-mass nucleus with moderate isospin asymmetry and a well-characterized structure, serving as a benchmark for the universal relation in a different mass region.
    \item[(c)] \text{$^{132}$Sn}: a doubly magic, neutron-rich nucleus with large isospin asymmetry, providing a stringent benchmark for testing EOS predictions in heavier systems strongly influenced by symmetry-energy effects.
\end{itemize}

Figure~\ref{fig:pred} shows $\zeta$ as a function of the electric dipole polarizability $\alpha_D$ for the three selected nuclei. The horizontal shaded bands indicate the CNSP-4 and CNSP-10 constraints, while the exponential curves correspond to the universal relation given in Eq.~\eqref{eq:un_rel} for each nucleus. The intersections of the horizontal constraints with the universal curves yield predictions for $\alpha_D$, which are highlighted by the vertical shaded bands. The corresponding values and uncertainties are summarized in Table~\ref{tab:table3}.

For comparison, we also overlay microscopic EDF-based calculations of $\alpha_D$ for each nucleus. In all cases, the predicted values fall within the range spanned by the microscopic model results. This agreement demonstrates that the universal relation not only captures the global trends across the calibrated functionals, but also exhibits predictive capability for neutron-rich nuclei not included in the fitting procedure.

\begin{table}
	\caption{Dipole polarizability values and their uncertainty for $^{52}$Ca, $^{90}$Zr, and $^{132}$Sn, as predicted from Eq.~\eqref{eq:un_rel}.}
	\begin{ruledtabular}
        \begin{tabular}{lrrr}
            & Nuclei & $\alpha_{D}~{\rm (fm^{3})}$ & $\sigma~{\rm (fm^{3})}$ \\
            \hline
            \multicolumn{4}{c}{\vspace{-0.25cm}} \\
            \multirow[t]{3}{*}{CNSP-4} & $^{52}$Ca & 2.911 & 0.084 \\
            & $^{90}$Zr & 5.423 & 0.133 \\
            & $^{132}$Sn & 9.816 & 0.231 \\
            \multirow[t]{3}{*}{CNSP-10} & $^{52}$Ca & 2.840 & 0.066 \\
            & $^{90}$Zr & 5.311 & 0.108 \\
            & $^{132}$Sn & 9.623 & 0.182 \\
        \end{tabular}
    \end{ruledtabular}
	\label{tab:table3}
\end{table}

%======================================================================================================================
\section{Conclusion}
\label{sec:conclusions}
%======================================================================================================================
This study establishes a universal relation bridging finite nuclei and neutron stars, in which the electric dipole polarizability $\alpha_D$ is linked to the compactness of $1.4~M_{\odot}$ neutron star, mediated by the slope of the symmetry energy. Both quantities serve as
complementary probes of the density dependence of the nuclear symmetry energy, with $\alpha_D$ reflecting nuclear structure and $\beta_{1.4}$ capturing neutron star properties. The emergence of this relation reveals a fundamental connection between microscopic nuclear properties and macroscopic neutron star structure, providing a robust, EOS-insensitive framework that unites laboratory measurements with astrophysical observations. This framework possesses a twofold predictive power: it enables informed predictions of nuclear observables in unexplored regions of the nuclear chart and, simultaneously, translates nuclear constraints into quantitative bounds on neutron star radii and the slope of the symmetry energy.

As experimental determinations of $\alpha_D$ extend to increasingly neutron-rich isotopes and astrophysical observations achieve higher precision, this framework can be further refined to yield tighter constraints on the density dependence of the symmetry energy and neutron star structure. In this way, the present work progress the connection between nuclear and neutron star matter, grounded in shared isovector properties of finite nuclei and supported by both theoretical modeling and experimental data, highlighting the link between terrestrial nuclei and one of the densest astrophysical objects in the universe.

%======================================================================================================================
\section*{Acknowledgments}
%======================================================================================================================
This work is supported by the Croatian Science Foundation under the project number HRZZ-MOBDOL-12-2023-6026 and under the project Relativistic Nuclear Many-Body Theory in the Multimessenger Observation Era (HRZZ-IP-2022-10-7773). E.Y. acknowledges support from the UK STFC under award no. ST/Y000358/1.
N.P. acknowledges support from the project “Implementation of cutting-edge research and its application as part of the Scientific Center of Excellence for Quantum and Complex Systems, and Representations of Lie Algebras”, Grant No. PK.1.1.10.0004, co-financed by the European Union through the European Regional Development Fund - Competitiveness and Cohesion Programme 2021-2027. This paper was supported by the European Union – NextGenerationEU through the National Recovery and Resilience Plan 2021-2026 -- Institutional grant of University of Zagreb Faculty of Science (Nuclear Astrophysics).

%======================================================================================================================
\section*{Author Contribution}
%======================================================================================================================
The conceptual framework of the study was developed by P.S.K. and N.P. Methodological design and computational implementation were carried out by P.S.K. and N.P., with P.S.K. responsible for model validation, formal analysis, visualization, and figure preparation. Model calculations and data collection and curation were performed by P.S.K., E.Y., and T.G. The manuscript was drafted by P.S.K., and revised by all authors. Supervision and project coordination were provided by N.P. Financial support for this work was secured by P.S.K., E.Y., and N.P.

%\nocite{*}
%======================================================================================================================
\bibliography{bibliography}% Produces the bibliography via BibTeX.

\begin{thebibliography}{112}%
\makeatletter
\providecommand \@ifxundefined [1]{%
 \@ifx{#1\undefined}
}%
\providecommand \@ifnum [1]{%
 \ifnum #1\expandafter \@firstoftwo
 \else \expandafter \@secondoftwo
 \fi
}%
\providecommand \@ifx [1]{%
 \ifx #1\expandafter \@firstoftwo
 \else \expandafter \@secondoftwo
 \fi
}%
\providecommand \natexlab [1]{#1}%
\providecommand \enquote  [1]{``#1''}%
\providecommand \bibnamefont  [1]{#1}%
\providecommand \bibfnamefont [1]{#1}%
\providecommand \citenamefont [1]{#1}%
\providecommand \href@noop [0]{\@secondoftwo}%
\providecommand \href [0]{\begingroup \@sanitize@url \@href}%
\providecommand \@href[1]{\@@startlink{#1}\@@href}%
\providecommand \@@href[1]{\endgroup#1\@@endlink}%
\providecommand \@sanitize@url [0]{\catcode `\\12\catcode `\$12\catcode `\&12\catcode `\#12\catcode `\^12\catcode `\_12\catcode `\%12\relax}%
\providecommand \@@startlink[1]{}%
\providecommand \@@endlink[0]{}%
\providecommand \url  [0]{\begingroup\@sanitize@url \@url }%
\providecommand \@url [1]{\endgroup\@href {#1}{\urlprefix }}%
\providecommand \urlprefix  [0]{URL }%
\providecommand \Eprint [0]{\href }%
\providecommand \doibase [0]{https://doi.org/}%
\providecommand \selectlanguage [0]{\@gobble}%
\providecommand \bibinfo  [0]{\@secondoftwo}%
\providecommand \bibfield  [0]{\@secondoftwo}%
\providecommand \translation [1]{[#1]}%
\providecommand \BibitemOpen [0]{}%
\providecommand \bibitemStop [0]{}%
\providecommand \bibitemNoStop [0]{.\EOS\space}%
\providecommand \EOS [0]{\spacefactor3000\relax}%
\providecommand \BibitemShut  [1]{\csname bibitem#1\endcsname}%
\let\auto@bib@innerbib\@empty
%</preamble>
\bibitem [{\citenamefont {Lattimer}\ and\ \citenamefont {Prakash}(2004)}]{Lattimer-2004}%
  \BibitemOpen
  \bibfield  {author} {\bibinfo {author} {\bibfnamefont {J.~M.}\ \bibnamefont {Lattimer}}\ and\ \bibinfo {author} {\bibfnamefont {M.}~\bibnamefont {Prakash}},\ }\bibfield  {title} {\bibinfo {title} {The physics of neutron stars},\ }\href {https://doi.org/10.1126/science.1090720} {\bibfield  {journal} {\bibinfo  {journal} {Sci.}\ }\textbf {\bibinfo {volume} {304}},\ \bibinfo {pages} {536} (\bibinfo {year} {2004})}\BibitemShut {NoStop}%
\bibitem [{\citenamefont {Lattimer}(2021)}]{Lattimer2021}%
  \BibitemOpen
  \bibfield  {author} {\bibinfo {author} {\bibfnamefont {J.}~\bibnamefont {Lattimer}},\ }\bibfield  {title} {\bibinfo {title} {Neutron {S}tars and the {N}uclear {M}atter {E}quation of {S}tate},\ }\href {https://doi.org/https://doi.org/10.1146/annurev-nucl-102419-124827} {\bibfield  {journal} {\bibinfo  {journal} {Ann. Rev. Nucl. Part. Sci.}\ }\textbf {\bibinfo {volume} {71}},\ \bibinfo {pages} {433} (\bibinfo {year} {2021})}\BibitemShut {NoStop}%
\bibitem [{\citenamefont {Burgio}\ \emph {et~al.}(2021)\citenamefont {Burgio}, \citenamefont {Schulze}, \citenamefont {Vidaña},\ and\ \citenamefont {Wei}}]{BURGIO2021103879}%
  \BibitemOpen
  \bibfield  {author} {\bibinfo {author} {\bibfnamefont {G.}~\bibnamefont {Burgio}}, \bibinfo {author} {\bibfnamefont {H.-J.}\ \bibnamefont {Schulze}}, \bibinfo {author} {\bibfnamefont {I.}~\bibnamefont {Vidaña}},\ and\ \bibinfo {author} {\bibfnamefont {J.-B.}\ \bibnamefont {Wei}},\ }\bibfield  {title} {\bibinfo {title} {Neutron stars and the nuclear equation of state},\ }\href {https://doi.org/https://doi.org/10.1016/j.ppnp.2021.103879} {\bibfield  {journal} {\bibinfo  {journal} {Prog. Part. Nucl. Phys.}\ }\textbf {\bibinfo {volume} {120}},\ \bibinfo {pages} {103879} (\bibinfo {year} {2021})}\BibitemShut {NoStop}%
\bibitem [{\citenamefont {Arzoumanian}\ \emph {et~al.}(2018)\citenamefont {Arzoumanian} \emph {et~al.}}]{Arzoumanian-2018}%
  \BibitemOpen
  \bibfield  {author} {\bibinfo {author} {\bibfnamefont {Z.}~\bibnamefont {Arzoumanian}} \emph {et~al.},\ }\bibfield  {title} {\bibinfo {title} {The {NANOG}rav 11-year {D}ata {S}et: {H}igh-precision {T}iming of 45 {M}illisecond {P}ulsars},\ }\href {https://doi.org/10.3847/1538-4365/aab5b0} {\bibfield  {journal} {\bibinfo  {journal} {Astrophys. J. Suppl. S.}\ }\textbf {\bibinfo {volume} {235}},\ \bibinfo {pages} {37} (\bibinfo {year} {2018})}\BibitemShut {NoStop}%
\bibitem [{\citenamefont {Antoniadis}\ \emph {et~al.}(2013)\citenamefont {Antoniadis} \emph {et~al.}}]{Antoniadis-2013}%
  \BibitemOpen
  \bibfield  {author} {\bibinfo {author} {\bibfnamefont {J.}~\bibnamefont {Antoniadis}} \emph {et~al.},\ }\bibfield  {title} {\bibinfo {title} {A {M}assive {P}ulsar in a {C}ompact {R}elativistic {B}inary},\ }\href {https://doi.org/10.1126/science.1233232} {\bibfield  {journal} {\bibinfo  {journal} {Sci.}\ }\textbf {\bibinfo {volume} {340}},\ \bibinfo {pages} {1233232} (\bibinfo {year} {2013})}\BibitemShut {NoStop}%
\bibitem [{\citenamefont {Cromartie}\ \emph {et~al.}(2020)\citenamefont {Cromartie} \emph {et~al.}}]{Cromartie-2020}%
  \BibitemOpen
  \bibfield  {author} {\bibinfo {author} {\bibfnamefont {H.}~\bibnamefont {Cromartie}} \emph {et~al.},\ }\bibfield  {title} {\bibinfo {title} {Relativistic {S}hapiro delay measurements of an extremely massive millisecond pulsar.},\ }\href {https://doi.org/https://doi.org/10.1038/s41550-019-0880-2} {\bibfield  {journal} {\bibinfo  {journal} {Nat. Astron.}\ }\textbf {\bibinfo {volume} {4}},\ \bibinfo {pages} {72} (\bibinfo {year} {2020})}\BibitemShut {NoStop}%
\bibitem [{\citenamefont {Fonseca}\ \emph {et~al.}(2021)\citenamefont {Fonseca} \emph {et~al.}}]{Fonseca_2021}%
  \BibitemOpen
  \bibfield  {author} {\bibinfo {author} {\bibfnamefont {E.}~\bibnamefont {Fonseca}} \emph {et~al.},\ }\bibfield  {title} {\bibinfo {title} {Refined {M}ass and {G}eometric {M}easurements of the {H}igh-mass {PSR} {J}0740+6620},\ }\href {https://doi.org/10.3847/2041-8213/ac03b8} {\bibfield  {journal} {\bibinfo  {journal} {Astrophys. J. Lett.}\ }\textbf {\bibinfo {volume} {915}},\ \bibinfo {pages} {L12} (\bibinfo {year} {2021})}\BibitemShut {NoStop}%
\bibitem [{\citenamefont {Romani}\ \emph {et~al.}(2022)\citenamefont {Romani}, \citenamefont {Kandel}, \citenamefont {Filippenko}, \citenamefont {Brink},\ and\ \citenamefont {Zheng}}]{Romani-2022}%
  \BibitemOpen
  \bibfield  {author} {\bibinfo {author} {\bibfnamefont {R.~W.}\ \bibnamefont {Romani}}, \bibinfo {author} {\bibfnamefont {D.}~\bibnamefont {Kandel}}, \bibinfo {author} {\bibfnamefont {A.~V.}\ \bibnamefont {Filippenko}}, \bibinfo {author} {\bibfnamefont {T.~G.}\ \bibnamefont {Brink}},\ and\ \bibinfo {author} {\bibfnamefont {W.}~\bibnamefont {Zheng}},\ }\bibfield  {title} {\bibinfo {title} {{PSR} {J}0952-0607: {T}he {F}astest and {H}eaviest {K}nown {G}alactic {N}eutron {S}tar},\ }\href {https://doi.org/10.3847/2041-8213/ac8007} {\bibfield  {journal} {\bibinfo  {journal} {Astrophys. J. Lett.}\ }\textbf {\bibinfo {volume} {934}},\ \bibinfo {pages} {L17} (\bibinfo {year} {2022})}\BibitemShut {NoStop}%
\bibitem [{\citenamefont {Miller}\ \emph {et~al.}(2019)\citenamefont {Miller} \emph {et~al.}}]{Miller_2019}%
  \BibitemOpen
  \bibfield  {author} {\bibinfo {author} {\bibfnamefont {M.~C.}\ \bibnamefont {Miller}} \emph {et~al.},\ }\bibfield  {title} {\bibinfo {title} {{PSR} {J}0030+0451 {M}ass and {R}adius from {NICER} {D}ata and {I}mplications for the {P}roperties of {N}eutron {S}tar {M}atter},\ }\href {https://doi.org/10.3847/2041-8213/ab50c5} {\bibfield  {journal} {\bibinfo  {journal} {Astrophys. J. Lett.}\ }\textbf {\bibinfo {volume} {887}},\ \bibinfo {pages} {L24} (\bibinfo {year} {2019})}\BibitemShut {NoStop}%
\bibitem [{\citenamefont {Riley}\ \emph {et~al.}(2019)\citenamefont {Riley} \emph {et~al.}}]{Riley_2019}%
  \BibitemOpen
  \bibfield  {author} {\bibinfo {author} {\bibfnamefont {T.~E.}\ \bibnamefont {Riley}} \emph {et~al.},\ }\bibfield  {title} {\bibinfo {title} {A {NICER} {V}iew of {PSR} {J}0030+0451: {M}illisecond {P}ulsar {P}arameter {E}stimation},\ }\href {https://doi.org/10.3847/2041-8213/ab481c} {\bibfield  {journal} {\bibinfo  {journal} {Astrophys. J. Lett.}\ }\textbf {\bibinfo {volume} {887}},\ \bibinfo {pages} {L21} (\bibinfo {year} {2019})}\BibitemShut {NoStop}%
\bibitem [{\citenamefont {Raaijmakers}\ \emph {et~al.}(2019)\citenamefont {Raaijmakers} \emph {et~al.}}]{Raaijmakers_2019}%
  \BibitemOpen
  \bibfield  {author} {\bibinfo {author} {\bibfnamefont {G.}~\bibnamefont {Raaijmakers}} \emph {et~al.},\ }\bibfield  {title} {\bibinfo {title} {A {NICER} {V}iew of {PSR} {J}0030+0451: {I}mplications for the {D}ense {M}atter {E}quation of {S}tate},\ }\href {https://doi.org/10.3847/2041-8213/ab451a} {\bibfield  {journal} {\bibinfo  {journal} {Astrophys. J. Lett.}\ }\textbf {\bibinfo {volume} {887}},\ \bibinfo {pages} {L22} (\bibinfo {year} {2019})}\BibitemShut {NoStop}%
\bibitem [{\citenamefont {Miller}\ \emph {et~al.}(2021)\citenamefont {Miller} \emph {et~al.}}]{Miller_2021}%
  \BibitemOpen
  \bibfield  {author} {\bibinfo {author} {\bibfnamefont {M.~C.}\ \bibnamefont {Miller}} \emph {et~al.},\ }\bibfield  {title} {\bibinfo {title} {{T}he {R}adius of {PSR} {J}0740+6620 from {NICER} and {XMM}-{N}ewton {D}ata},\ }\href {https://doi.org/10.3847/2041-8213/ac089b} {\bibfield  {journal} {\bibinfo  {journal} {Astrophys. J. Lett.}\ }\textbf {\bibinfo {volume} {918}},\ \bibinfo {pages} {L28} (\bibinfo {year} {2021})}\BibitemShut {NoStop}%
\bibitem [{\citenamefont {Riley}\ \emph {et~al.}(2021)\citenamefont {Riley} \emph {et~al.}}]{Riley_2021}%
  \BibitemOpen
  \bibfield  {author} {\bibinfo {author} {\bibfnamefont {T.~E.}\ \bibnamefont {Riley}} \emph {et~al.},\ }\bibfield  {title} {\bibinfo {title} {{A} {NICER} {V}iew of the {M}assive {P}ulsar {PSR} {J}0740+6620 {I}nformed by {R}adio {T}iming and {XMM}-{N}ewton {S}pectroscopy},\ }\href {https://doi.org/10.3847/2041-8213/ac0a81} {\bibfield  {journal} {\bibinfo  {journal} {Astrophys. J. Lett.}\ }\textbf {\bibinfo {volume} {918}},\ \bibinfo {pages} {L27} (\bibinfo {year} {2021})}\BibitemShut {NoStop}%
\bibitem [{\citenamefont {Dittmann}\ \emph {et~al.}(2024)\citenamefont {Dittmann} \emph {et~al.}}]{Dittmann_2024}%
  \BibitemOpen
  \bibfield  {author} {\bibinfo {author} {\bibfnamefont {A.~J.}\ \bibnamefont {Dittmann}} \emph {et~al.},\ }\bibfield  {title} {\bibinfo {title} {A {M}ore {P}recise {M}easurement of the {R}adius of {PSR} {J}0740+6620 {U}sing {U}pdated {NICER} {D}ata},\ }\href {https://doi.org/10.3847/1538-4357/ad5f1e} {\bibfield  {journal} {\bibinfo  {journal} {Astrophys. J.}\ }\textbf {\bibinfo {volume} {974}},\ \bibinfo {pages} {295} (\bibinfo {year} {2024})}\BibitemShut {NoStop}%
\bibitem [{\citenamefont {Abbott}\ \emph {et~al.}(2019)\citenamefont {Abbott} \emph {et~al.}}]{Abbott-2019}%
  \BibitemOpen
  \bibfield  {author} {\bibinfo {author} {\bibfnamefont {B.~P.}\ \bibnamefont {Abbott}} \emph {et~al.} (\bibinfo {collaboration} {LIGO Scientific Collaboration and Virgo Collaboration}),\ }\bibfield  {title} {\bibinfo {title} {Properties of the {B}inary {N}eutron {S}tar {M}erger {GW170817}},\ }\href {https://doi.org/10.1103/PhysRevX.9.011001} {\bibfield  {journal} {\bibinfo  {journal} {Phys. Rev. X}\ }\textbf {\bibinfo {volume} {9}},\ \bibinfo {pages} {011001} (\bibinfo {year} {2019})}\BibitemShut {NoStop}%
\bibitem [{\citenamefont {Abbott}\ \emph {et~al.}(2017)\citenamefont {Abbott} \emph {et~al.}}]{PhysRevLett.119.161101}%
  \BibitemOpen
  \bibfield  {author} {\bibinfo {author} {\bibfnamefont {B.~P.}\ \bibnamefont {Abbott}} \emph {et~al.} (\bibinfo {collaboration} {LIGO Scientific Collaboration and Virgo Collaboration}),\ }\bibfield  {title} {\bibinfo {title} {{GW}170817: Observation of gravitational waves from a binary neutron star inspiral},\ }\href {https://doi.org/10.1103/PhysRevLett.119.161101} {\bibfield  {journal} {\bibinfo  {journal} {Phys. Rev. Lett.}\ }\textbf {\bibinfo {volume} {119}},\ \bibinfo {pages} {161101} (\bibinfo {year} {2017})}\BibitemShut {NoStop}%
\bibitem [{\citenamefont {Abbott}\ \emph {et~al.}(2020)\citenamefont {Abbott} \emph {et~al.}}]{Abbott_2020}%
  \BibitemOpen
  \bibfield  {author} {\bibinfo {author} {\bibfnamefont {B.~P.}\ \bibnamefont {Abbott}} \emph {et~al.},\ }\bibfield  {title} {\bibinfo {title} {{GW}190425: {O}bservation of a {C}ompact {B}inary {C}oalescence with {T}otal {M}ass $\sim$ 3.4 {M}$_{\odot}$},\ }\href {https://doi.org/10.3847/2041-8213/ab75f5} {\bibfield  {journal} {\bibinfo  {journal} {Astrophys. J. Lett.}\ }\textbf {\bibinfo {volume} {892}},\ \bibinfo {pages} {L3} (\bibinfo {year} {2020})}\BibitemShut {NoStop}%
\bibitem [{\citenamefont {Hessels}\ \emph {et~al.}(2006)\citenamefont {Hessels}, \citenamefont {Ransom}, \citenamefont {Stairs}, \citenamefont {Freire}, \citenamefont {Kaspi},\ and\ \citenamefont {Camilo}}]{doi:10.1126/science.1123430}%
  \BibitemOpen
  \bibfield  {author} {\bibinfo {author} {\bibfnamefont {J.~W.~T.}\ \bibnamefont {Hessels}}, \bibinfo {author} {\bibfnamefont {S.~M.}\ \bibnamefont {Ransom}}, \bibinfo {author} {\bibfnamefont {I.~H.}\ \bibnamefont {Stairs}}, \bibinfo {author} {\bibfnamefont {P.~C.~C.}\ \bibnamefont {Freire}}, \bibinfo {author} {\bibfnamefont {V.~M.}\ \bibnamefont {Kaspi}},\ and\ \bibinfo {author} {\bibfnamefont {F.}~\bibnamefont {Camilo}},\ }\bibfield  {title} {\bibinfo {title} {A {R}adio {P}ulsar {S}pinning at 716 {H}z},\ }\href {https://doi.org/10.1126/science.1123430} {\bibfield  {journal} {\bibinfo  {journal} {Sci.}\ }\textbf {\bibinfo {volume} {311}},\ \bibinfo {pages} {1901} (\bibinfo {year} {2006})}\BibitemShut {NoStop}%
\bibitem [{\citenamefont {Bejger}\ \emph {et~al.}(2005)\citenamefont {Bejger}, \citenamefont {Bulik},\ and\ \citenamefont {Haensel}}]{10.1111/j.1365-2966.2005.09575.x}%
  \BibitemOpen
  \bibfield  {author} {\bibinfo {author} {\bibfnamefont {M.}~\bibnamefont {Bejger}}, \bibinfo {author} {\bibfnamefont {T.}~\bibnamefont {Bulik}},\ and\ \bibinfo {author} {\bibfnamefont {P.}~\bibnamefont {Haensel}},\ }\bibfield  {title} {\bibinfo {title} {Constraints on the dense matter equation of state from the measurements of {PSR} {J}0737-3039{A} moment of inertia and {PSR} {J}0751+1807 mass},\ }\href {https://doi.org/10.1111/j.1365-2966.2005.09575.x} {\bibfield  {journal} {\bibinfo  {journal} {Mon. Not. Roy. Astron. Soc.}\ }\textbf {\bibinfo {volume} {364}},\ \bibinfo {pages} {635} (\bibinfo {year} {2005})}\BibitemShut {NoStop}%
\bibitem [{\citenamefont {Koliogiannis}\ and\ \citenamefont {Moustakidis}(2020)}]{PhysRevC.101.015805}%
  \BibitemOpen
  \bibfield  {author} {\bibinfo {author} {\bibfnamefont {P.~S.}\ \bibnamefont {Koliogiannis}}\ and\ \bibinfo {author} {\bibfnamefont {C.~C.}\ \bibnamefont {Moustakidis}},\ }\bibfield  {title} {\bibinfo {title} {Effects of the equation of state on the bulk properties of maximally rotating neutron stars},\ }\href {https://doi.org/10.1103/PhysRevC.101.015805} {\bibfield  {journal} {\bibinfo  {journal} {Phys. Rev. C}\ }\textbf {\bibinfo {volume} {101}},\ \bibinfo {pages} {015805} (\bibinfo {year} {2020})}\BibitemShut {NoStop}%
\bibitem [{\citenamefont {Roca-Maza}\ and\ \citenamefont {Paar}(2018)}]{ROCAMAZA201896}%
  \BibitemOpen
  \bibfield  {author} {\bibinfo {author} {\bibfnamefont {X.}~\bibnamefont {Roca-Maza}}\ and\ \bibinfo {author} {\bibfnamefont {N.}~\bibnamefont {Paar}},\ }\bibfield  {title} {\bibinfo {title} {Nuclear equation of state from ground and collective excited state properties of nuclei},\ }\href {https://doi.org/https://doi.org/10.1016/j.ppnp.2018.04.001} {\bibfield  {journal} {\bibinfo  {journal} {Prog. Part. Nucl. Phys.}\ }\textbf {\bibinfo {volume} {101}},\ \bibinfo {pages} {96} (\bibinfo {year} {2018})}\BibitemShut {NoStop}%
\bibitem [{\citenamefont {Birkhan}\ \emph {et~al.}(2017)\citenamefont {Birkhan} \emph {et~al.}}]{Birkhan2017}%
  \BibitemOpen
  \bibfield  {author} {\bibinfo {author} {\bibfnamefont {J.}~\bibnamefont {Birkhan}} \emph {et~al.},\ }\bibfield  {title} {\bibinfo {title} {Electric {D}ipole {P}olarizability of $^{48}\mathrm{Ca}$ and {I}mplications for the {N}eutron {S}kin},\ }\href {https://doi.org/10.1103/PhysRevLett.118.252501} {\bibfield  {journal} {\bibinfo  {journal} {Phys. Rev. Lett.}\ }\textbf {\bibinfo {volume} {118}},\ \bibinfo {pages} {252501} (\bibinfo {year} {2017})}\BibitemShut {NoStop}%
\bibitem [{\citenamefont {Fattoyev}\ \emph {et~al.}(2018)\citenamefont {Fattoyev}, \citenamefont {Piekarewicz},\ and\ \citenamefont {Horowitz}}]{PhysRevLett.120.172702}%
  \BibitemOpen
  \bibfield  {author} {\bibinfo {author} {\bibfnamefont {F.~J.}\ \bibnamefont {Fattoyev}}, \bibinfo {author} {\bibfnamefont {J.}~\bibnamefont {Piekarewicz}},\ and\ \bibinfo {author} {\bibfnamefont {C.~J.}\ \bibnamefont {Horowitz}},\ }\bibfield  {title} {\bibinfo {title} {Neutron {S}kins and {N}eutron {S}tars in the {M}ultimessenger {E}ra},\ }\href {https://doi.org/10.1103/PhysRevLett.120.172702} {\bibfield  {journal} {\bibinfo  {journal} {Phys. Rev. Lett.}\ }\textbf {\bibinfo {volume} {120}},\ \bibinfo {pages} {172702} (\bibinfo {year} {2018})}\BibitemShut {NoStop}%
\bibitem [{\citenamefont {Rossi}\ \emph {et~al.}(2013)\citenamefont {Rossi} \emph {et~al.}}]{PhysRevLett.111.242503}%
  \BibitemOpen
  \bibfield  {author} {\bibinfo {author} {\bibfnamefont {D.~M.}\ \bibnamefont {Rossi}} \emph {et~al.},\ }\bibfield  {title} {\bibinfo {title} {Measurement of the {D}ipole {P}olarizability of the {U}nstable {N}eutron-{R}ich {N}ucleus $^{68}\mathrm{Ni}$},\ }\href {https://doi.org/10.1103/PhysRevLett.111.242503} {\bibfield  {journal} {\bibinfo  {journal} {Phys. Rev. Lett.}\ }\textbf {\bibinfo {volume} {111}},\ \bibinfo {pages} {242503} (\bibinfo {year} {2013})}\BibitemShut {NoStop}%
\bibitem [{\citenamefont {Roca-Maza}\ \emph {et~al.}(2015)\citenamefont {Roca-Maza}, \citenamefont {Vi\~nas}, \citenamefont {Centelles}, \citenamefont {Agrawal}, \citenamefont {Col\`o}, \citenamefont {Paar}, \citenamefont {Piekarewicz},\ and\ \citenamefont {Vretenar}}]{Rokamaza2015}%
  \BibitemOpen
  \bibfield  {author} {\bibinfo {author} {\bibfnamefont {X.}~\bibnamefont {Roca-Maza}}, \bibinfo {author} {\bibfnamefont {X.}~\bibnamefont {Vi\~nas}}, \bibinfo {author} {\bibfnamefont {M.}~\bibnamefont {Centelles}}, \bibinfo {author} {\bibfnamefont {B.~K.}\ \bibnamefont {Agrawal}}, \bibinfo {author} {\bibfnamefont {G.}~\bibnamefont {Col\`o}}, \bibinfo {author} {\bibfnamefont {N.}~\bibnamefont {Paar}}, \bibinfo {author} {\bibfnamefont {J.}~\bibnamefont {Piekarewicz}},\ and\ \bibinfo {author} {\bibfnamefont {D.}~\bibnamefont {Vretenar}},\ }\bibfield  {title} {\bibinfo {title} {Neutron skin thickness from the measured electric dipole polarizability in $^{68}\text{Ni}$, $^{120}\text{Sn}$, and $^{208}\text{Pb}$},\ }\href {https://doi.org/10.1103/PhysRevC.92.064304} {\bibfield  {journal} {\bibinfo  {journal} {Phys. Rev. C}\ }\textbf {\bibinfo {volume} {92}},\ \bibinfo {pages} {064304} (\bibinfo {year} {2015})}\BibitemShut {NoStop}%
\bibitem [{\citenamefont {Tamii}\ \emph {et~al.}(2011)\citenamefont {Tamii} \emph {et~al.}}]{PhysRevLett.107.062502}%
  \BibitemOpen
  \bibfield  {author} {\bibinfo {author} {\bibfnamefont {A.}~\bibnamefont {Tamii}} \emph {et~al.},\ }\bibfield  {title} {\bibinfo {title} {Complete {E}lectric {D}ipole {R}esponse and the {N}eutron {S}kin in $^{208}\mathrm{Pb}$},\ }\href {https://doi.org/10.1103/PhysRevLett.107.062502} {\bibfield  {journal} {\bibinfo  {journal} {Phys. Rev. Lett.}\ }\textbf {\bibinfo {volume} {107}},\ \bibinfo {pages} {062502} (\bibinfo {year} {2011})}\BibitemShut {NoStop}%
\bibitem [{\citenamefont {Piekarewicz}\ \emph {et~al.}(2012)\citenamefont {Piekarewicz}, \citenamefont {Agrawal}, \citenamefont {Col\`o}, \citenamefont {Nazarewicz}, \citenamefont {Paar}, \citenamefont {Reinhard}, \citenamefont {Roca-Maza},\ and\ \citenamefont {Vretenar}}]{PhysRevC.85.041302}%
  \BibitemOpen
  \bibfield  {author} {\bibinfo {author} {\bibfnamefont {J.}~\bibnamefont {Piekarewicz}}, \bibinfo {author} {\bibfnamefont {B.~K.}\ \bibnamefont {Agrawal}}, \bibinfo {author} {\bibfnamefont {G.}~\bibnamefont {Col\`o}}, \bibinfo {author} {\bibfnamefont {W.}~\bibnamefont {Nazarewicz}}, \bibinfo {author} {\bibfnamefont {N.}~\bibnamefont {Paar}}, \bibinfo {author} {\bibfnamefont {P.-G.}\ \bibnamefont {Reinhard}}, \bibinfo {author} {\bibfnamefont {X.}~\bibnamefont {Roca-Maza}},\ and\ \bibinfo {author} {\bibfnamefont {D.}~\bibnamefont {Vretenar}},\ }\bibfield  {title} {\bibinfo {title} {Electric dipole polarizability and the neutron skin},\ }\href {https://doi.org/10.1103/PhysRevC.85.041302} {\bibfield  {journal} {\bibinfo  {journal} {Phys. Rev. C}\ }\textbf {\bibinfo {volume} {85}},\ \bibinfo {pages} {041302} (\bibinfo {year} {2012})}\BibitemShut {NoStop}%
\bibitem [{\citenamefont {Hashimoto}\ \emph {et~al.}(2015)\citenamefont {Hashimoto} \emph {et~al.}}]{PhysRevC.92.031305}%
  \BibitemOpen
  \bibfield  {author} {\bibinfo {author} {\bibfnamefont {T.}~\bibnamefont {Hashimoto}} \emph {et~al.},\ }\bibfield  {title} {\bibinfo {title} {Dipole polarizability of $^{120}\mathrm{Sn}$ and nuclear energy density functionals},\ }\href {https://doi.org/10.1103/PhysRevC.92.031305} {\bibfield  {journal} {\bibinfo  {journal} {Phys. Rev. C}\ }\textbf {\bibinfo {volume} {92}},\ \bibinfo {pages} {031305} (\bibinfo {year} {2015})}\BibitemShut {NoStop}%
\bibitem [{\citenamefont {Bassauer}\ \emph {et~al.}(2020)\citenamefont {Bassauer} \emph {et~al.}}]{Bassauer2020}%
  \BibitemOpen
  \bibfield  {author} {\bibinfo {author} {\bibfnamefont {S.}~\bibnamefont {Bassauer}} \emph {et~al.},\ }\bibfield  {title} {\bibinfo {title} {Evolution of the dipole polarizability in the stable tin isotope chain},\ }\href {https://doi.org/https://doi.org/10.1016/j.physletb.2020.135804} {\bibfield  {journal} {\bibinfo  {journal} {Phys. Lett. B}\ }\textbf {\bibinfo {volume} {810}},\ \bibinfo {pages} {135804} (\bibinfo {year} {2020})}\BibitemShut {NoStop}%
\bibitem [{\citenamefont {Brandherm}\ \emph {et~al.}(2025)\citenamefont {Brandherm} \emph {et~al.}}]{PhysRevC.111.024312}%
  \BibitemOpen
  \bibfield  {author} {\bibinfo {author} {\bibfnamefont {I.}~\bibnamefont {Brandherm}} \emph {et~al.},\ }\bibfield  {title} {\bibinfo {title} {Electric dipole polarizability of $^{58}\mathrm{Ni}$},\ }\href {https://doi.org/10.1103/PhysRevC.111.024312} {\bibfield  {journal} {\bibinfo  {journal} {Phys. Rev. C}\ }\textbf {\bibinfo {volume} {111}},\ \bibinfo {pages} {024312} (\bibinfo {year} {2025})}\BibitemShut {NoStop}%
\bibitem [{\citenamefont {Zhang}\ and\ \citenamefont {Chen}(2015)}]{PhysRevC.92.031301}%
  \BibitemOpen
  \bibfield  {author} {\bibinfo {author} {\bibfnamefont {Z.}~\bibnamefont {Zhang}}\ and\ \bibinfo {author} {\bibfnamefont {L.-W.}\ \bibnamefont {Chen}},\ }\bibfield  {title} {\bibinfo {title} {Electric dipole polarizability in $^{208}\mathbf{Pb}$ as a probe of the symmetry energy and neutron matter around ${\ensuremath{\rho}}_{0}/3$},\ }\href {https://doi.org/10.1103/PhysRevC.92.031301} {\bibfield  {journal} {\bibinfo  {journal} {Phys. Rev. C}\ }\textbf {\bibinfo {volume} {92}},\ \bibinfo {pages} {031301} (\bibinfo {year} {2015})}\BibitemShut {NoStop}%
\bibitem [{\citenamefont {Russotto}\ \emph {et~al.}(2016)\citenamefont {Russotto} \emph {et~al.}}]{PhysRevC.94.034608}%
  \BibitemOpen
  \bibfield  {author} {\bibinfo {author} {\bibfnamefont {P.}~\bibnamefont {Russotto}} \emph {et~al.},\ }\bibfield  {title} {\bibinfo {title} {Results of the {ASY-EOS} experiment at {GSI}: {T}he symmetry energy at suprasaturation density},\ }\href {https://doi.org/10.1103/PhysRevC.94.034608} {\bibfield  {journal} {\bibinfo  {journal} {Phys. Rev. C}\ }\textbf {\bibinfo {volume} {94}},\ \bibinfo {pages} {034608} (\bibinfo {year} {2016})}\BibitemShut {NoStop}%
\bibitem [{\citenamefont {{Le Fèvre}}\ \emph {et~al.}(2016)\citenamefont {{Le Fèvre}}, \citenamefont {Leifels}, \citenamefont {Reisdorf}, \citenamefont {Aichelin},\ and\ \citenamefont {Hartnack}}]{LEFEVRE2016112}%
  \BibitemOpen
  \bibfield  {author} {\bibinfo {author} {\bibfnamefont {A.}~\bibnamefont {{Le Fèvre}}}, \bibinfo {author} {\bibfnamefont {Y.}~\bibnamefont {Leifels}}, \bibinfo {author} {\bibfnamefont {W.}~\bibnamefont {Reisdorf}}, \bibinfo {author} {\bibfnamefont {J.}~\bibnamefont {Aichelin}},\ and\ \bibinfo {author} {\bibfnamefont {C.}~\bibnamefont {Hartnack}},\ }\bibfield  {title} {\bibinfo {title} {Constraining the nuclear matter equation of state around twice saturation density},\ }\href {https://doi.org/https://doi.org/10.1016/j.nuclphysa.2015.09.015} {\bibfield  {journal} {\bibinfo  {journal} {Nucl. Phys. A}\ }\textbf {\bibinfo {volume} {945}},\ \bibinfo {pages} {112} (\bibinfo {year} {2016})}\BibitemShut {NoStop}%
\bibitem [{\citenamefont {Huth}\ \emph {et~al.}(2022)\citenamefont {Huth} \emph {et~al.}}]{Huth_2022}%
  \BibitemOpen
  \bibfield  {author} {\bibinfo {author} {\bibfnamefont {S.}~\bibnamefont {Huth}} \emph {et~al.},\ }\bibfield  {title} {\bibinfo {title} {Constraining neutron-star matter with microscopic and macroscopic collisions},\ }\href {https://doi.org/10.1038/s41586-022-04750-w} {\bibfield  {journal} {\bibinfo  {journal} {Nat.}\ }\textbf {\bibinfo {volume} {606}},\ \bibinfo {pages} {276} (\bibinfo {year} {2022})}\BibitemShut {NoStop}%
\bibitem [{\citenamefont {Yagi}\ and\ \citenamefont {Yunes}(2013{\natexlab{a}})}]{doi:10.1126/science.1236462}%
  \BibitemOpen
  \bibfield  {author} {\bibinfo {author} {\bibfnamefont {K.}~\bibnamefont {Yagi}}\ and\ \bibinfo {author} {\bibfnamefont {N.}~\bibnamefont {Yunes}},\ }\bibfield  {title} {\bibinfo {title} {{I}-{L}ove-{Q}: {U}nexpected {U}niversal {R}elations for {N}eutron {S}tars and {Q}uark {S}tars},\ }\href {https://doi.org/10.1126/science.1236462} {\bibfield  {journal} {\bibinfo  {journal} {Sci.}\ }\textbf {\bibinfo {volume} {341}},\ \bibinfo {pages} {365} (\bibinfo {year} {2013}{\natexlab{a}})}\BibitemShut {NoStop}%
\bibitem [{\citenamefont {Maselli}\ \emph {et~al.}(2013)\citenamefont {Maselli}, \citenamefont {Cardoso}, \citenamefont {Ferrari}, \citenamefont {Gualtieri},\ and\ \citenamefont {Pani}}]{PhysRevD.88.023007}%
  \BibitemOpen
  \bibfield  {author} {\bibinfo {author} {\bibfnamefont {A.}~\bibnamefont {Maselli}}, \bibinfo {author} {\bibfnamefont {V.}~\bibnamefont {Cardoso}}, \bibinfo {author} {\bibfnamefont {V.}~\bibnamefont {Ferrari}}, \bibinfo {author} {\bibfnamefont {L.}~\bibnamefont {Gualtieri}},\ and\ \bibinfo {author} {\bibfnamefont {P.}~\bibnamefont {Pani}},\ }\bibfield  {title} {\bibinfo {title} {Equation-of-state-independent relations in neutron stars},\ }\href {https://doi.org/10.1103/PhysRevD.88.023007} {\bibfield  {journal} {\bibinfo  {journal} {Phys. Rev. D}\ }\textbf {\bibinfo {volume} {88}},\ \bibinfo {pages} {023007} (\bibinfo {year} {2013})}\BibitemShut {NoStop}%
\bibitem [{\citenamefont {Yagi}\ and\ \citenamefont {Yunes}(2013{\natexlab{b}})}]{PhysRevD.88.023009}%
  \BibitemOpen
  \bibfield  {author} {\bibinfo {author} {\bibfnamefont {K.}~\bibnamefont {Yagi}}\ and\ \bibinfo {author} {\bibfnamefont {N.}~\bibnamefont {Yunes}},\ }\bibfield  {title} {\bibinfo {title} {{I}-{L}ove-{Q} relations in neutron stars and their applications to astrophysics, gravitational waves, and fundamental physics},\ }\href {https://doi.org/10.1103/PhysRevD.88.023009} {\bibfield  {journal} {\bibinfo  {journal} {Phys. Rev. D}\ }\textbf {\bibinfo {volume} {88}},\ \bibinfo {pages} {023009} (\bibinfo {year} {2013}{\natexlab{b}})}\BibitemShut {NoStop}%
\bibitem [{\citenamefont {Chakrabarti}\ \emph {et~al.}(2014)\citenamefont {Chakrabarti}, \citenamefont {Delsate}, \citenamefont {G\"urlebeck},\ and\ \citenamefont {Steinhoff}}]{PhysRevLett.112.201102}%
  \BibitemOpen
  \bibfield  {author} {\bibinfo {author} {\bibfnamefont {S.}~\bibnamefont {Chakrabarti}}, \bibinfo {author} {\bibfnamefont {T.}~\bibnamefont {Delsate}}, \bibinfo {author} {\bibfnamefont {N.}~\bibnamefont {G\"urlebeck}},\ and\ \bibinfo {author} {\bibfnamefont {J.}~\bibnamefont {Steinhoff}},\ }\bibfield  {title} {\bibinfo {title} {{I}$\text{\ensuremath{-}}${Q} {R}elation for {R}apidly {R}otating {N}eutron {S}tars},\ }\href {https://doi.org/10.1103/PhysRevLett.112.201102} {\bibfield  {journal} {\bibinfo  {journal} {Phys. Rev. Lett.}\ }\textbf {\bibinfo {volume} {112}},\ \bibinfo {pages} {201102} (\bibinfo {year} {2014})}\BibitemShut {NoStop}%
\bibitem [{\citenamefont {Jiang}\ and\ \citenamefont {Yagi}(2020)}]{PhysRevD.101.124006}%
  \BibitemOpen
  \bibfield  {author} {\bibinfo {author} {\bibfnamefont {N.}~\bibnamefont {Jiang}}\ and\ \bibinfo {author} {\bibfnamefont {K.}~\bibnamefont {Yagi}},\ }\bibfield  {title} {\bibinfo {title} {Analytic {I}-{L}ove-{C} relations for realistic neutron stars},\ }\href {https://doi.org/10.1103/PhysRevD.101.124006} {\bibfield  {journal} {\bibinfo  {journal} {Phys. Rev. D}\ }\textbf {\bibinfo {volume} {101}},\ \bibinfo {pages} {124006} (\bibinfo {year} {2020})}\BibitemShut {NoStop}%
\bibitem [{\citenamefont {Lowrey}\ \emph {et~al.}(2025)\citenamefont {Lowrey}, \citenamefont {Yagi},\ and\ \citenamefont {Yunes}}]{PhysRevD.111.024075}%
  \BibitemOpen
  \bibfield  {author} {\bibinfo {author} {\bibfnamefont {T.}~\bibnamefont {Lowrey}}, \bibinfo {author} {\bibfnamefont {K.}~\bibnamefont {Yagi}},\ and\ \bibinfo {author} {\bibfnamefont {N.}~\bibnamefont {Yunes}},\ }\bibfield  {title} {\bibinfo {title} {Improved analytic {L}ove-{C} relations for neutron stars},\ }\href {https://doi.org/10.1103/PhysRevD.111.024075} {\bibfield  {journal} {\bibinfo  {journal} {Phys. Rev. D}\ }\textbf {\bibinfo {volume} {111}},\ \bibinfo {pages} {024075} (\bibinfo {year} {2025})}\BibitemShut {NoStop}%
\bibitem [{\citenamefont {Lattimer}\ and\ \citenamefont {Prakash}(2001)}]{Lattimer_2001}%
  \BibitemOpen
  \bibfield  {author} {\bibinfo {author} {\bibfnamefont {J.~M.}\ \bibnamefont {Lattimer}}\ and\ \bibinfo {author} {\bibfnamefont {M.}~\bibnamefont {Prakash}},\ }\bibfield  {title} {\bibinfo {title} {Neutron {S}tar {S}tructure and the {E}quation of {S}tate},\ }\href {https://doi.org/10.1086/319702} {\bibfield  {journal} {\bibinfo  {journal} {Astrophys. J.}\ }\textbf {\bibinfo {volume} {550}},\ \bibinfo {pages} {426} (\bibinfo {year} {2001})}\BibitemShut {NoStop}%
\bibitem [{\citenamefont {Reed}\ and\ \citenamefont {Horowitz}(2020)}]{PhysRevD.102.103011}%
  \BibitemOpen
  \bibfield  {author} {\bibinfo {author} {\bibfnamefont {B.}~\bibnamefont {Reed}}\ and\ \bibinfo {author} {\bibfnamefont {C.~J.}\ \bibnamefont {Horowitz}},\ }\bibfield  {title} {\bibinfo {title} {Total energy in supernova neutrinos and the tidal deformability and binding energy of neutron stars},\ }\href {https://doi.org/10.1103/PhysRevD.102.103011} {\bibfield  {journal} {\bibinfo  {journal} {Phys. Rev. D}\ }\textbf {\bibinfo {volume} {102}},\ \bibinfo {pages} {103011} (\bibinfo {year} {2020})}\BibitemShut {NoStop}%
\bibitem [{\citenamefont {Laskos-Patkos}\ \emph {et~al.}(2022)\citenamefont {Laskos-Patkos}, \citenamefont {Koliogiannis}, \citenamefont {Kanakis-Pegios},\ and\ \citenamefont {Moustakidis}}]{universe8080395}%
  \BibitemOpen
  \bibfield  {author} {\bibinfo {author} {\bibfnamefont {P.}~\bibnamefont {Laskos-Patkos}}, \bibinfo {author} {\bibfnamefont {P.~S.}\ \bibnamefont {Koliogiannis}}, \bibinfo {author} {\bibfnamefont {A.}~\bibnamefont {Kanakis-Pegios}},\ and\ \bibinfo {author} {\bibfnamefont {C.~C.}\ \bibnamefont {Moustakidis}},\ }\bibfield  {title} {\bibinfo {title} {{T}hermodynamics of {H}ot {N}eutron {S}tars and {U}niversal {R}elations},\ }\href {https://doi.org/10.3390/universe8080395} {\bibfield  {journal} {\bibinfo  {journal} {Universe}\ }\textbf {\bibinfo {volume} {8}},\ \bibinfo {pages} {395} (\bibinfo {year} {2022})}\BibitemShut {NoStop}%
\bibitem [{\citenamefont {Lattimer}\ and\ \citenamefont {Prakash}(2007)}]{LATTIMER2007109}%
  \BibitemOpen
  \bibfield  {author} {\bibinfo {author} {\bibfnamefont {J.~M.}\ \bibnamefont {Lattimer}}\ and\ \bibinfo {author} {\bibfnamefont {M.}~\bibnamefont {Prakash}},\ }\bibfield  {title} {\bibinfo {title} {Neutron star observations: Prognosis for equation of state constraints},\ }\href {https://doi.org/https://doi.org/10.1016/j.physrep.2007.02.003} {\bibfield  {journal} {\bibinfo  {journal} {Phys. Rep.}\ }\textbf {\bibinfo {volume} {442}},\ \bibinfo {pages} {109} (\bibinfo {year} {2007})},\ \bibinfo {note} {the Hans Bethe Centennial Volume 1906-2006}\BibitemShut {NoStop}%
\bibitem [{\citenamefont {Doneva}\ \emph {et~al.}(2013{\natexlab{a}})\citenamefont {Doneva}, \citenamefont {Yazadjiev}, \citenamefont {Stergioulas},\ and\ \citenamefont {Kokkotas}}]{Doneva_2014}%
  \BibitemOpen
  \bibfield  {author} {\bibinfo {author} {\bibfnamefont {D.~D.}\ \bibnamefont {Doneva}}, \bibinfo {author} {\bibfnamefont {S.~S.}\ \bibnamefont {Yazadjiev}}, \bibinfo {author} {\bibfnamefont {N.}~\bibnamefont {Stergioulas}},\ and\ \bibinfo {author} {\bibfnamefont {K.~D.}\ \bibnamefont {Kokkotas}},\ }\bibfield  {title} {\bibinfo {title} {Breakdown of {I}–{L}ove–{Q} {U}niversality in {R}apidly {R}otating {R}elativistic {S}tars},\ }\href {https://doi.org/10.1088/2041-8205/781/1/L6} {\bibfield  {journal} {\bibinfo  {journal} {Astrophys. J. Lett.}\ }\textbf {\bibinfo {volume} {781}},\ \bibinfo {pages} {L6} (\bibinfo {year} {2013}{\natexlab{a}})}\BibitemShut {NoStop}%
\bibitem [{\citenamefont {Martinon}\ \emph {et~al.}(2014)\citenamefont {Martinon}, \citenamefont {Maselli}, \citenamefont {Gualtieri},\ and\ \citenamefont {Ferrari}}]{PhysRevD.90.064026}%
  \BibitemOpen
  \bibfield  {author} {\bibinfo {author} {\bibfnamefont {G.}~\bibnamefont {Martinon}}, \bibinfo {author} {\bibfnamefont {A.}~\bibnamefont {Maselli}}, \bibinfo {author} {\bibfnamefont {L.}~\bibnamefont {Gualtieri}},\ and\ \bibinfo {author} {\bibfnamefont {V.}~\bibnamefont {Ferrari}},\ }\bibfield  {title} {\bibinfo {title} {Rotating protoneutron stars: Spin evolution, maximum mass, and {I}-{L}ove-{Q} relations},\ }\href {https://doi.org/10.1103/PhysRevD.90.064026} {\bibfield  {journal} {\bibinfo  {journal} {Phys. Rev. D}\ }\textbf {\bibinfo {volume} {90}},\ \bibinfo {pages} {064026} (\bibinfo {year} {2014})}\BibitemShut {NoStop}%
\bibitem [{\citenamefont {Breu}\ and\ \citenamefont {Rezzolla}(2016)}]{Breu_2016}%
  \BibitemOpen
  \bibfield  {author} {\bibinfo {author} {\bibfnamefont {C.}~\bibnamefont {Breu}}\ and\ \bibinfo {author} {\bibfnamefont {L.}~\bibnamefont {Rezzolla}},\ }\bibfield  {title} {\bibinfo {title} {Maximum mass, moment of inertia and compactness of relativistic stars},\ }\href {https://doi.org/10.1093/mnras/stw575} {\bibfield  {journal} {\bibinfo  {journal} {Mon. Not. Roy. Astron. Soc.}\ }\textbf {\bibinfo {volume} {459}},\ \bibinfo {pages} {646} (\bibinfo {year} {2016})}\BibitemShut {NoStop}%
\bibitem [{\citenamefont {Cipolletta}\ \emph {et~al.}(2017)\citenamefont {Cipolletta}, \citenamefont {Cherubini}, \citenamefont {Filippi}, \citenamefont {Rueda},\ and\ \citenamefont {Ruffini}}]{PhysRevD.96.024046}%
  \BibitemOpen
  \bibfield  {author} {\bibinfo {author} {\bibfnamefont {F.}~\bibnamefont {Cipolletta}}, \bibinfo {author} {\bibfnamefont {C.}~\bibnamefont {Cherubini}}, \bibinfo {author} {\bibfnamefont {S.}~\bibnamefont {Filippi}}, \bibinfo {author} {\bibfnamefont {J.~A.}\ \bibnamefont {Rueda}},\ and\ \bibinfo {author} {\bibfnamefont {R.}~\bibnamefont {Ruffini}},\ }\bibfield  {title} {\bibinfo {title} {Last stable orbit around rapidly rotating neutron stars},\ }\href {https://doi.org/10.1103/PhysRevD.96.024046} {\bibfield  {journal} {\bibinfo  {journal} {Phys. Rev. D}\ }\textbf {\bibinfo {volume} {96}},\ \bibinfo {pages} {024046} (\bibinfo {year} {2017})}\BibitemShut {NoStop}%
\bibitem [{\citenamefont {Luk}\ and\ \citenamefont {Lin}(2018)}]{Luk_2018}%
  \BibitemOpen
  \bibfield  {author} {\bibinfo {author} {\bibfnamefont {S.-S.}\ \bibnamefont {Luk}}\ and\ \bibinfo {author} {\bibfnamefont {L.-M.}\ \bibnamefont {Lin}},\ }\bibfield  {title} {\bibinfo {title} {Universal {R}elations for {I}nnermost {S}table {C}ircular {O}rbits around {R}apidly {R}otating {N}eutron {S}tars},\ }\href {https://doi.org/10.3847/1538-4357/aac8d6} {\bibfield  {journal} {\bibinfo  {journal} {Astrophys. J.}\ }\textbf {\bibinfo {volume} {861}},\ \bibinfo {pages} {141} (\bibinfo {year} {2018})}\BibitemShut {NoStop}%
\bibitem [{\citenamefont {Riahi}\ \emph {et~al.}(2019)\citenamefont {Riahi}, \citenamefont {Kalantari},\ and\ \citenamefont {Rueda}}]{PhysRevD.99.043004}%
  \BibitemOpen
  \bibfield  {author} {\bibinfo {author} {\bibfnamefont {R.}~\bibnamefont {Riahi}}, \bibinfo {author} {\bibfnamefont {S.~Z.}\ \bibnamefont {Kalantari}},\ and\ \bibinfo {author} {\bibfnamefont {J.~A.}\ \bibnamefont {Rueda}},\ }\bibfield  {title} {\bibinfo {title} {Universal relations for the keplerian sequence of rotating neutron stars},\ }\href {https://doi.org/10.1103/PhysRevD.99.043004} {\bibfield  {journal} {\bibinfo  {journal} {Phys. Rev. D}\ }\textbf {\bibinfo {volume} {99}},\ \bibinfo {pages} {043004} (\bibinfo {year} {2019})}\BibitemShut {NoStop}%
\bibitem [{\citenamefont {Sun}\ \emph {et~al.}(2020)\citenamefont {Sun}, \citenamefont {Wen},\ and\ \citenamefont {Wang}}]{PhysRevD.102.023039}%
  \BibitemOpen
  \bibfield  {author} {\bibinfo {author} {\bibfnamefont {W.}~\bibnamefont {Sun}}, \bibinfo {author} {\bibfnamefont {D.}~\bibnamefont {Wen}},\ and\ \bibinfo {author} {\bibfnamefont {J.}~\bibnamefont {Wang}},\ }\bibfield  {title} {\bibinfo {title} {New quasiuniversal relations for static and rapid rotating neutron stars},\ }\href {https://doi.org/10.1103/PhysRevD.102.023039} {\bibfield  {journal} {\bibinfo  {journal} {Phys. Rev. D}\ }\textbf {\bibinfo {volume} {102}},\ \bibinfo {pages} {023039} (\bibinfo {year} {2020})}\BibitemShut {NoStop}%
\bibitem [{\citenamefont {Silva}\ \emph {et~al.}(2021)\citenamefont {Silva}, \citenamefont {Pappas}, \citenamefont {Yunes},\ and\ \citenamefont {Yagi}}]{PhysRevD.103.063038}%
  \BibitemOpen
  \bibfield  {author} {\bibinfo {author} {\bibfnamefont {H.~O.}\ \bibnamefont {Silva}}, \bibinfo {author} {\bibfnamefont {G.}~\bibnamefont {Pappas}}, \bibinfo {author} {\bibfnamefont {N.}~\bibnamefont {Yunes}},\ and\ \bibinfo {author} {\bibfnamefont {K.}~\bibnamefont {Yagi}},\ }\bibfield  {title} {\bibinfo {title} {Surface of rapidly-rotating neutron stars: {I}mplications to neutron star parameter estimation},\ }\href {https://doi.org/10.1103/PhysRevD.103.063038} {\bibfield  {journal} {\bibinfo  {journal} {Phys. Rev. D}\ }\textbf {\bibinfo {volume} {103}},\ \bibinfo {pages} {063038} (\bibinfo {year} {2021})}\BibitemShut {NoStop}%
\bibitem [{\citenamefont {Koliogiannis}\ and\ \citenamefont {Moustakidis}(2021)}]{Koliogiannis_2021}%
  \BibitemOpen
  \bibfield  {author} {\bibinfo {author} {\bibfnamefont {P.~S.}\ \bibnamefont {Koliogiannis}}\ and\ \bibinfo {author} {\bibfnamefont {C.~C.}\ \bibnamefont {Moustakidis}},\ }\bibfield  {title} {\bibinfo {title} {Thermodynamical description of hot, rapidly rotating neutron stars, protoneutron stars, and neutron star merger remnants},\ }\href {https://doi.org/10.3847/1538-4357/abe542} {\bibfield  {journal} {\bibinfo  {journal} {Astrophys. J.}\ }\textbf {\bibinfo {volume} {912}},\ \bibinfo {pages} {69} (\bibinfo {year} {2021})}\BibitemShut {NoStop}%
\bibitem [{\citenamefont {Papigkiotis}\ and\ \citenamefont {Pappas}(2023)}]{PhysRevD.107.103050}%
  \BibitemOpen
  \bibfield  {author} {\bibinfo {author} {\bibfnamefont {G.}~\bibnamefont {Papigkiotis}}\ and\ \bibinfo {author} {\bibfnamefont {G.}~\bibnamefont {Pappas}},\ }\bibfield  {title} {\bibinfo {title} {Universal relations for rapidly rotating neutron stars using supervised machine-learning techniques},\ }\href {https://doi.org/10.1103/PhysRevD.107.103050} {\bibfield  {journal} {\bibinfo  {journal} {Phys. Rev. D}\ }\textbf {\bibinfo {volume} {107}},\ \bibinfo {pages} {103050} (\bibinfo {year} {2023})}\BibitemShut {NoStop}%
\bibitem [{\citenamefont {Musolino}\ \emph {et~al.}(2024)\citenamefont {Musolino}, \citenamefont {Ecker},\ and\ \citenamefont {Rezzolla}}]{Musolino_2024}%
  \BibitemOpen
  \bibfield  {author} {\bibinfo {author} {\bibfnamefont {C.}~\bibnamefont {Musolino}}, \bibinfo {author} {\bibfnamefont {C.}~\bibnamefont {Ecker}},\ and\ \bibinfo {author} {\bibfnamefont {L.}~\bibnamefont {Rezzolla}},\ }\bibfield  {title} {\bibinfo {title} {On the {M}aximum {M}ass and {O}blateness of {R}otating {N}eutron {S}tars with {G}eneric {E}quations of {S}tate},\ }\href {https://doi.org/10.3847/1538-4357/ad1758} {\bibfield  {journal} {\bibinfo  {journal} {Astrophys. J.}\ }\textbf {\bibinfo {volume} {962}},\ \bibinfo {pages} {61} (\bibinfo {year} {2024})}\BibitemShut {NoStop}%
\bibitem [{\citenamefont {Manoharan}\ and\ \citenamefont {Kokkotas}(2024)}]{PhysRevD.109.103033}%
  \BibitemOpen
  \bibfield  {author} {\bibinfo {author} {\bibfnamefont {P.}~\bibnamefont {Manoharan}}\ and\ \bibinfo {author} {\bibfnamefont {K.~D.}\ \bibnamefont {Kokkotas}},\ }\bibfield  {title} {\bibinfo {title} {Finding universal relations using statistical data analysis},\ }\href {https://doi.org/10.1103/PhysRevD.109.103033} {\bibfield  {journal} {\bibinfo  {journal} {Phys. Rev. D}\ }\textbf {\bibinfo {volume} {109}},\ \bibinfo {pages} {103033} (\bibinfo {year} {2024})}\BibitemShut {NoStop}%
\bibitem [{\citenamefont {Marques}\ \emph {et~al.}(2017)\citenamefont {Marques}, \citenamefont {Oertel}, \citenamefont {Hempel},\ and\ \citenamefont {Novak}}]{PhysRevC.96.045806}%
  \BibitemOpen
  \bibfield  {author} {\bibinfo {author} {\bibfnamefont {M.}~\bibnamefont {Marques}}, \bibinfo {author} {\bibfnamefont {M.}~\bibnamefont {Oertel}}, \bibinfo {author} {\bibfnamefont {M.}~\bibnamefont {Hempel}},\ and\ \bibinfo {author} {\bibfnamefont {J.}~\bibnamefont {Novak}},\ }\bibfield  {title} {\bibinfo {title} {New temperature dependent hyperonic equation of state: {A}pplication to rotating neutron star models and {I}$\text{\ensuremath{-}}${Q} relations},\ }\href {https://doi.org/10.1103/PhysRevC.96.045806} {\bibfield  {journal} {\bibinfo  {journal} {Phys. Rev. C}\ }\textbf {\bibinfo {volume} {96}},\ \bibinfo {pages} {045806} (\bibinfo {year} {2017})}\BibitemShut {NoStop}%
\bibitem [{\citenamefont {Raduta}\ \emph {et~al.}(2020)\citenamefont {Raduta}, \citenamefont {Oertel},\ and\ \citenamefont {Sedrakian}}]{Raduta_2020}%
  \BibitemOpen
  \bibfield  {author} {\bibinfo {author} {\bibfnamefont {A.~R.}\ \bibnamefont {Raduta}}, \bibinfo {author} {\bibfnamefont {M.}~\bibnamefont {Oertel}},\ and\ \bibinfo {author} {\bibfnamefont {A.}~\bibnamefont {Sedrakian}},\ }\bibfield  {title} {\bibinfo {title} {Proto-neutron stars with heavy baryons and universal relations},\ }\href {https://doi.org/10.1093/mnras/staa2491} {\bibfield  {journal} {\bibinfo  {journal} {Mon. Not. Roy. Astron. Soc.}\ }\textbf {\bibinfo {volume} {499}},\ \bibinfo {pages} {914} (\bibinfo {year} {2020})}\BibitemShut {NoStop}%
\bibitem [{\citenamefont {Khadkikar}\ \emph {et~al.}(2021)\citenamefont {Khadkikar}, \citenamefont {Raduta}, \citenamefont {Oertel},\ and\ \citenamefont {Sedrakian}}]{PhysRevC.103.055811}%
  \BibitemOpen
  \bibfield  {author} {\bibinfo {author} {\bibfnamefont {S.}~\bibnamefont {Khadkikar}}, \bibinfo {author} {\bibfnamefont {A.~R.}\ \bibnamefont {Raduta}}, \bibinfo {author} {\bibfnamefont {M.}~\bibnamefont {Oertel}},\ and\ \bibinfo {author} {\bibfnamefont {A.}~\bibnamefont {Sedrakian}},\ }\bibfield  {title} {\bibinfo {title} {Maximum mass of compact stars from gravitational wave events with finite-temperature equations of state},\ }\href {https://doi.org/10.1103/PhysRevC.103.055811} {\bibfield  {journal} {\bibinfo  {journal} {Phys. Rev. C}\ }\textbf {\bibinfo {volume} {103}},\ \bibinfo {pages} {055811} (\bibinfo {year} {2021})}\BibitemShut {NoStop}%
\bibitem [{\citenamefont {Yagi}\ and\ \citenamefont {Yunes}(2017)}]{YAGI20171}%
  \BibitemOpen
  \bibfield  {author} {\bibinfo {author} {\bibfnamefont {K.}~\bibnamefont {Yagi}}\ and\ \bibinfo {author} {\bibfnamefont {N.}~\bibnamefont {Yunes}},\ }\bibfield  {title} {\bibinfo {title} {Approximate universal relations for neutron stars and quark stars},\ }\href {https://doi.org/https://doi.org/10.1016/j.physrep.2017.03.002} {\bibfield  {journal} {\bibinfo  {journal} {Phys. Rep.}\ }\textbf {\bibinfo {volume} {681}},\ \bibinfo {pages} {1} (\bibinfo {year} {2017})}\BibitemShut {NoStop}%
\bibitem [{\citenamefont {Wei}\ \emph {et~al.}(2019)\citenamefont {Wei}, \citenamefont {Figura}, \citenamefont {Burgio}, \citenamefont {Chen},\ and\ \citenamefont {Schulze}}]{Wei_2019}%
  \BibitemOpen
  \bibfield  {author} {\bibinfo {author} {\bibfnamefont {J.-B.}\ \bibnamefont {Wei}}, \bibinfo {author} {\bibfnamefont {A.}~\bibnamefont {Figura}}, \bibinfo {author} {\bibfnamefont {G.~F.}\ \bibnamefont {Burgio}}, \bibinfo {author} {\bibfnamefont {H.}~\bibnamefont {Chen}},\ and\ \bibinfo {author} {\bibfnamefont {H.-J.}\ \bibnamefont {Schulze}},\ }\bibfield  {title} {\bibinfo {title} {Neutron star universal relations with microscopic equations of state},\ }\href {https://doi.org/10.1088/1361-6471/aaf95c} {\bibfield  {journal} {\bibinfo  {journal} {J. Phys. G: Nucl. Part. Phys.}\ }\textbf {\bibinfo {volume} {46}},\ \bibinfo {pages} {034001} (\bibinfo {year} {2019})}\BibitemShut {NoStop}%
\bibitem [{\citenamefont {Khosravi~Largani}\ \emph {et~al.}(2022)\citenamefont {Khosravi~Largani}, \citenamefont {Fischer}, \citenamefont {Sedrakian}, \citenamefont {Cierniak}, \citenamefont {Alvarez-Castillo},\ and\ \citenamefont {Blaschke}}]{Khosravi_2022}%
  \BibitemOpen
  \bibfield  {author} {\bibinfo {author} {\bibfnamefont {N.}~\bibnamefont {Khosravi~Largani}}, \bibinfo {author} {\bibfnamefont {T.}~\bibnamefont {Fischer}}, \bibinfo {author} {\bibfnamefont {A.}~\bibnamefont {Sedrakian}}, \bibinfo {author} {\bibfnamefont {M.}~\bibnamefont {Cierniak}}, \bibinfo {author} {\bibfnamefont {D.~E.}\ \bibnamefont {Alvarez-Castillo}},\ and\ \bibinfo {author} {\bibfnamefont {D.~B.}\ \bibnamefont {Blaschke}},\ }\bibfield  {title} {\bibinfo {title} {Universal relations for rapidly rotating cold and hot hybrid stars},\ }\href {https://doi.org/10.1093/mnras/stac1916} {\bibfield  {journal} {\bibinfo  {journal} {Mon. Not. Roy. Astron. Soc.}\ }\textbf {\bibinfo {volume} {515}},\ \bibinfo {pages} {3539} (\bibinfo {year} {2022})}\BibitemShut {NoStop}%
\bibitem [{\citenamefont {Kumar}\ \emph {et~al.}(2024)\citenamefont {Kumar}, \citenamefont {Ghosh}, \citenamefont {Thakur}, \citenamefont {Thapa}, \citenamefont {Nath},\ and\ \citenamefont {Sinha}}]{Kumar_2024}%
  \BibitemOpen
  \bibfield  {author} {\bibinfo {author} {\bibfnamefont {A.}~\bibnamefont {Kumar}}, \bibinfo {author} {\bibfnamefont {M.~K.}\ \bibnamefont {Ghosh}}, \bibinfo {author} {\bibfnamefont {P.}~\bibnamefont {Thakur}}, \bibinfo {author} {\bibfnamefont {V.~B.}\ \bibnamefont {Thapa}}, \bibinfo {author} {\bibfnamefont {K.~K.}\ \bibnamefont {Nath}},\ and\ \bibinfo {author} {\bibfnamefont {M.}~\bibnamefont {Sinha}},\ }\bibfield  {title} {\bibinfo {title} {Universal relations for compact stars with exotic degrees of freedom},\ }\href {https://doi.org/10.1140/epjc/s10052-024-13066-0} {\bibfield  {journal} {\bibinfo  {journal} {Eur. Phys. J. C}\ }\textbf {\bibinfo {volume} {84}},\ \bibinfo {pages} {692} (\bibinfo {year} {2024})}\BibitemShut {NoStop}%
\bibitem [{\citenamefont {Koliogiannis}\ and\ \citenamefont {Moustakidis}(2019)}]{Koliogiannis_2019}%
  \BibitemOpen
  \bibfield  {author} {\bibinfo {author} {\bibfnamefont {P.~S.}\ \bibnamefont {Koliogiannis}}\ and\ \bibinfo {author} {\bibfnamefont {C.~C.}\ \bibnamefont {Moustakidis}},\ }\bibfield  {title} {\bibinfo {title} {Constraints on the equation of state from the stability condition of neutron stars},\ }\href {https://doi.org/10.1007/s10509-019-3539-7} {\bibfield  {journal} {\bibinfo  {journal} {Astrophys. Space Sci.}\ }\textbf {\bibinfo {volume} {364}},\ \bibinfo {pages} {52} (\bibinfo {year} {2019})}\BibitemShut {NoStop}%
\bibitem [{\citenamefont {Andersson}\ and\ \citenamefont {Kokkotas}(1996)}]{PhysRevLett.77.4134}%
  \BibitemOpen
  \bibfield  {author} {\bibinfo {author} {\bibfnamefont {N.}~\bibnamefont {Andersson}}\ and\ \bibinfo {author} {\bibfnamefont {K.~D.}\ \bibnamefont {Kokkotas}},\ }\bibfield  {title} {\bibinfo {title} {Gravitational {W}aves and {P}ulsating stars: {W}hat {C}an {W}e {L}earn from {F}uture {O}bservations?},\ }\href {https://doi.org/10.1103/PhysRevLett.77.4134} {\bibfield  {journal} {\bibinfo  {journal} {Phys. Rev. Lett.}\ }\textbf {\bibinfo {volume} {77}},\ \bibinfo {pages} {4134} (\bibinfo {year} {1996})}\BibitemShut {NoStop}%
\bibitem [{\citenamefont {Andersson}\ and\ \citenamefont {Kokkotas}(1998)}]{10.1046/j.1365-8711.1998.01840.x}%
  \BibitemOpen
  \bibfield  {author} {\bibinfo {author} {\bibfnamefont {N.}~\bibnamefont {Andersson}}\ and\ \bibinfo {author} {\bibfnamefont {K.~D.}\ \bibnamefont {Kokkotas}},\ }\bibfield  {title} {\bibinfo {title} {Towards gravitational wave asteroseismology},\ }\href {https://doi.org/10.1046/j.1365-8711.1998.01840.x} {\bibfield  {journal} {\bibinfo  {journal} {Mon. Not. Roy. Astron. Soc.}\ }\textbf {\bibinfo {volume} {299}},\ \bibinfo {pages} {1059} (\bibinfo {year} {1998})}\BibitemShut {NoStop}%
\bibitem [{\citenamefont {Doneva}\ \emph {et~al.}(2013{\natexlab{b}})\citenamefont {Doneva}, \citenamefont {Gaertig}, \citenamefont {Kokkotas},\ and\ \citenamefont {Kr\"uger}}]{PhysRevD.88.044052}%
  \BibitemOpen
  \bibfield  {author} {\bibinfo {author} {\bibfnamefont {D.~D.}\ \bibnamefont {Doneva}}, \bibinfo {author} {\bibfnamefont {E.}~\bibnamefont {Gaertig}}, \bibinfo {author} {\bibfnamefont {K.~D.}\ \bibnamefont {Kokkotas}},\ and\ \bibinfo {author} {\bibfnamefont {C.}~\bibnamefont {Kr\"uger}},\ }\bibfield  {title} {\bibinfo {title} {Gravitational wave asteroseismology of fast rotating neutron stars with realistic equations of state},\ }\href {https://doi.org/10.1103/PhysRevD.88.044052} {\bibfield  {journal} {\bibinfo  {journal} {Phys. Rev. D}\ }\textbf {\bibinfo {volume} {88}},\ \bibinfo {pages} {044052} (\bibinfo {year} {2013}{\natexlab{b}})}\BibitemShut {NoStop}%
\bibitem [{\citenamefont {Lioutas}\ \emph {et~al.}(2021)\citenamefont {Lioutas}, \citenamefont {Bauswein},\ and\ \citenamefont {Stergioulas}}]{PhysRevD.104.043011}%
  \BibitemOpen
  \bibfield  {author} {\bibinfo {author} {\bibfnamefont {G.}~\bibnamefont {Lioutas}}, \bibinfo {author} {\bibfnamefont {A.}~\bibnamefont {Bauswein}},\ and\ \bibinfo {author} {\bibfnamefont {N.}~\bibnamefont {Stergioulas}},\ }\bibfield  {title} {\bibinfo {title} {Frequency deviations in universal relations of isolated neutron stars and postmerger remnants},\ }\href {https://doi.org/10.1103/PhysRevD.104.043011} {\bibfield  {journal} {\bibinfo  {journal} {Phys. Rev. D}\ }\textbf {\bibinfo {volume} {104}},\ \bibinfo {pages} {043011} (\bibinfo {year} {2021})}\BibitemShut {NoStop}%
\bibitem [{\citenamefont {Yagi}\ and\ \citenamefont {Yunes}(2016)}]{Yagi_2017}%
  \BibitemOpen
  \bibfield  {author} {\bibinfo {author} {\bibfnamefont {K.}~\bibnamefont {Yagi}}\ and\ \bibinfo {author} {\bibfnamefont {N.}~\bibnamefont {Yunes}},\ }\bibfield  {title} {\bibinfo {title} {Approximate universal relations among tidal parameters for neutron star binaries},\ }\href {https://doi.org/10.1088/1361-6382/34/1/015006} {\bibfield  {journal} {\bibinfo  {journal} {Class. Quantum Grav.}\ }\textbf {\bibinfo {volume} {34}},\ \bibinfo {pages} {015006} (\bibinfo {year} {2016})}\BibitemShut {NoStop}%
\bibitem [{\citenamefont {Kumar}\ and\ \citenamefont {Landry}(2019)}]{PhysRevD.99.123026}%
  \BibitemOpen
  \bibfield  {author} {\bibinfo {author} {\bibfnamefont {B.}~\bibnamefont {Kumar}}\ and\ \bibinfo {author} {\bibfnamefont {P.}~\bibnamefont {Landry}},\ }\bibfield  {title} {\bibinfo {title} {Inferring neutron star properties from {GW}170817 with universal relations},\ }\href {https://doi.org/10.1103/PhysRevD.99.123026} {\bibfield  {journal} {\bibinfo  {journal} {Phys. Rev. D}\ }\textbf {\bibinfo {volume} {99}},\ \bibinfo {pages} {123026} (\bibinfo {year} {2019})}\BibitemShut {NoStop}%
\bibitem [{\citenamefont {Saes}\ and\ \citenamefont {Mendes}(2022)}]{PhysRevD.106.043027}%
  \BibitemOpen
  \bibfield  {author} {\bibinfo {author} {\bibfnamefont {J.~A.}\ \bibnamefont {Saes}}\ and\ \bibinfo {author} {\bibfnamefont {R.~F.~P.}\ \bibnamefont {Mendes}},\ }\bibfield  {title} {\bibinfo {title} {Equation-of-state-insensitive measure of neutron star stiffness},\ }\href {https://doi.org/10.1103/PhysRevD.106.043027} {\bibfield  {journal} {\bibinfo  {journal} {Phys. Rev. D}\ }\textbf {\bibinfo {volume} {106}},\ \bibinfo {pages} {043027} (\bibinfo {year} {2022})}\BibitemShut {NoStop}%
\bibitem [{\citenamefont {Nath}\ \emph {et~al.}(2023)\citenamefont {Nath}, \citenamefont {Mallick},\ and\ \citenamefont {Chatterjee}}]{10.1093/mnras/stad1967}%
  \BibitemOpen
  \bibfield  {author} {\bibinfo {author} {\bibfnamefont {K.~K.}\ \bibnamefont {Nath}}, \bibinfo {author} {\bibfnamefont {R.}~\bibnamefont {Mallick}},\ and\ \bibinfo {author} {\bibfnamefont {S.}~\bibnamefont {Chatterjee}},\ }\bibfield  {title} {\bibinfo {title} {{I}-{L}ove-{Q} relations for a generic family of neutron star equations of state},\ }\href {https://doi.org/10.1093/mnras/stad1967} {\bibfield  {journal} {\bibinfo  {journal} {Mon. Not. Roy. Astron. Soc.}\ }\textbf {\bibinfo {volume} {524}},\ \bibinfo {pages} {1438} (\bibinfo {year} {2023})}\BibitemShut {NoStop}%
\bibitem [{\citenamefont {Pradhan}\ \emph {et~al.}(2023)\citenamefont {Pradhan}, \citenamefont {Vijaykumar},\ and\ \citenamefont {Chatterjee}}]{PhysRevD.107.023010}%
  \BibitemOpen
  \bibfield  {author} {\bibinfo {author} {\bibfnamefont {B.~K.}\ \bibnamefont {Pradhan}}, \bibinfo {author} {\bibfnamefont {A.}~\bibnamefont {Vijaykumar}},\ and\ \bibinfo {author} {\bibfnamefont {D.}~\bibnamefont {Chatterjee}},\ }\bibfield  {title} {\bibinfo {title} {Impact of updated multipole love numbers and $f$-{L}ove universal relations in the context of binary neutron stars},\ }\href {https://doi.org/10.1103/PhysRevD.107.023010} {\bibfield  {journal} {\bibinfo  {journal} {Phys. Rev. D}\ }\textbf {\bibinfo {volume} {107}},\ \bibinfo {pages} {023010} (\bibinfo {year} {2023})}\BibitemShut {NoStop}%
\bibitem [{\citenamefont {Aranguren}\ \emph {et~al.}(2023)\citenamefont {Aranguren}, \citenamefont {Font}, \citenamefont {Sanchis-Gual},\ and\ \citenamefont {Vera}}]{PhysRevD.108.104065}%
  \BibitemOpen
  \bibfield  {author} {\bibinfo {author} {\bibfnamefont {E.}~\bibnamefont {Aranguren}}, \bibinfo {author} {\bibfnamefont {J.~A.}\ \bibnamefont {Font}}, \bibinfo {author} {\bibfnamefont {N.}~\bibnamefont {Sanchis-Gual}},\ and\ \bibinfo {author} {\bibfnamefont {R.}~\bibnamefont {Vera}},\ }\bibfield  {title} {\bibinfo {title} {Revisiting the {$I$}-{L}ove-{$Q$} relations for superfluid neutron stars},\ }\href {https://doi.org/10.1103/PhysRevD.108.104065} {\bibfield  {journal} {\bibinfo  {journal} {Phys. Rev. D}\ }\textbf {\bibinfo {volume} {108}},\ \bibinfo {pages} {104065} (\bibinfo {year} {2023})}\BibitemShut {NoStop}%
\bibitem [{\citenamefont {Saes}\ \emph {et~al.}(2024)\citenamefont {Saes}, \citenamefont {Mendes},\ and\ \citenamefont {Yunes}}]{PhysRevD.110.024011}%
  \BibitemOpen
  \bibfield  {author} {\bibinfo {author} {\bibfnamefont {J.~A.}\ \bibnamefont {Saes}}, \bibinfo {author} {\bibfnamefont {R.~F.~P.}\ \bibnamefont {Mendes}},\ and\ \bibinfo {author} {\bibfnamefont {N.}~\bibnamefont {Yunes}},\ }\bibfield  {title} {\bibinfo {title} {Approximately universal {I}-{L}ove-$\langle{c}_{s}^{2}\rangle$ relations for the average neutron star stiffness},\ }\href {https://doi.org/10.1103/PhysRevD.110.024011} {\bibfield  {journal} {\bibinfo  {journal} {Phys. Rev. D}\ }\textbf {\bibinfo {volume} {110}},\ \bibinfo {pages} {024011} (\bibinfo {year} {2024})}\BibitemShut {NoStop}%
\bibitem [{\citenamefont {Suleiman}\ and\ \citenamefont {Read}(2024)}]{PhysRevD.109.103029}%
  \BibitemOpen
  \bibfield  {author} {\bibinfo {author} {\bibfnamefont {L.}~\bibnamefont {Suleiman}}\ and\ \bibinfo {author} {\bibfnamefont {J.}~\bibnamefont {Read}},\ }\bibfield  {title} {\bibinfo {title} {Quasiuniversal relations in the context of future neutron star detections},\ }\href {https://doi.org/10.1103/PhysRevD.109.103029} {\bibfield  {journal} {\bibinfo  {journal} {Phys. Rev. D}\ }\textbf {\bibinfo {volume} {109}},\ \bibinfo {pages} {103029} (\bibinfo {year} {2024})}\BibitemShut {NoStop}%
\bibitem [{\citenamefont {Legred}\ \emph {et~al.}(2024)\citenamefont {Legred}, \citenamefont {Sy-Garcia}, \citenamefont {Chatziioannou},\ and\ \citenamefont {Essick}}]{PhysRevD.109.023020}%
  \BibitemOpen
  \bibfield  {author} {\bibinfo {author} {\bibfnamefont {I.}~\bibnamefont {Legred}}, \bibinfo {author} {\bibfnamefont {B.~O.}\ \bibnamefont {Sy-Garcia}}, \bibinfo {author} {\bibfnamefont {K.}~\bibnamefont {Chatziioannou}},\ and\ \bibinfo {author} {\bibfnamefont {R.}~\bibnamefont {Essick}},\ }\bibfield  {title} {\bibinfo {title} {Assessing equation of state-independent relations for neutron stars with nonparametric models},\ }\href {https://doi.org/10.1103/PhysRevD.109.023020} {\bibfield  {journal} {\bibinfo  {journal} {Phys. Rev. D}\ }\textbf {\bibinfo {volume} {109}},\ \bibinfo {pages} {023020} (\bibinfo {year} {2024})}\BibitemShut {NoStop}%
\bibitem [{\citenamefont {Chatterjee}\ and\ \citenamefont {Nath}(2025)}]{Chatterjee_2025}%
  \BibitemOpen
  \bibfield  {author} {\bibinfo {author} {\bibfnamefont {S.}~\bibnamefont {Chatterjee}}\ and\ \bibinfo {author} {\bibfnamefont {K.~K.}\ \bibnamefont {Nath}},\ }\bibfield  {title} {\bibinfo {title} {Insights into neutron stars from gravitational redshifts and universal relations},\ }\href {https://doi.org/10.1140/epjc/s10052-025-14611-1} {\bibfield  {journal} {\bibinfo  {journal} {Eur. Phys. J. C}\ }\textbf {\bibinfo {volume} {85}},\ \bibinfo {pages} {862} (\bibinfo {year} {2025})}\BibitemShut {NoStop}%
\bibitem [{\citenamefont {Manoharan}\ \emph {et~al.}(2021)\citenamefont {Manoharan}, \citenamefont {Kr\"uger},\ and\ \citenamefont {Kokkotas}}]{PhysRevD.104.023005}%
  \BibitemOpen
  \bibfield  {author} {\bibinfo {author} {\bibfnamefont {P.}~\bibnamefont {Manoharan}}, \bibinfo {author} {\bibfnamefont {C.~J.}\ \bibnamefont {Kr\"uger}},\ and\ \bibinfo {author} {\bibfnamefont {K.~D.}\ \bibnamefont {Kokkotas}},\ }\bibfield  {title} {\bibinfo {title} {Universal relations for binary neutron star mergers with long-lived remnants},\ }\href {https://doi.org/10.1103/PhysRevD.104.023005} {\bibfield  {journal} {\bibinfo  {journal} {Phys. Rev. D}\ }\textbf {\bibinfo {volume} {104}},\ \bibinfo {pages} {023005} (\bibinfo {year} {2021})}\BibitemShut {NoStop}%
\bibitem [{\citenamefont {Godzieba}\ \emph {et~al.}(2021)\citenamefont {Godzieba}, \citenamefont {Gamba}, \citenamefont {Radice},\ and\ \citenamefont {Bernuzzi}}]{PhysRevD.103.063036}%
  \BibitemOpen
  \bibfield  {author} {\bibinfo {author} {\bibfnamefont {D.~A.}\ \bibnamefont {Godzieba}}, \bibinfo {author} {\bibfnamefont {R.}~\bibnamefont {Gamba}}, \bibinfo {author} {\bibfnamefont {D.}~\bibnamefont {Radice}},\ and\ \bibinfo {author} {\bibfnamefont {S.}~\bibnamefont {Bernuzzi}},\ }\bibfield  {title} {\bibinfo {title} {Updated universal relations for tidal deformabilities of neutron stars from phenomenological equations of state},\ }\href {https://doi.org/10.1103/PhysRevD.103.063036} {\bibfield  {journal} {\bibinfo  {journal} {Phys. Rev. D}\ }\textbf {\bibinfo {volume} {103}},\ \bibinfo {pages} {063036} (\bibinfo {year} {2021})}\BibitemShut {NoStop}%
\bibitem [{\citenamefont {Xie}\ \emph {et~al.}(2023)\citenamefont {Xie}, \citenamefont {Chatterjee}, \citenamefont {Holder}, \citenamefont {Holz}, \citenamefont {Perkins}, \citenamefont {Yagi},\ and\ \citenamefont {Yunes}}]{PhysRevD.107.043010}%
  \BibitemOpen
  \bibfield  {author} {\bibinfo {author} {\bibfnamefont {Y.}~\bibnamefont {Xie}}, \bibinfo {author} {\bibfnamefont {D.}~\bibnamefont {Chatterjee}}, \bibinfo {author} {\bibfnamefont {G.}~\bibnamefont {Holder}}, \bibinfo {author} {\bibfnamefont {D.~E.}\ \bibnamefont {Holz}}, \bibinfo {author} {\bibfnamefont {S.}~\bibnamefont {Perkins}}, \bibinfo {author} {\bibfnamefont {K.}~\bibnamefont {Yagi}},\ and\ \bibinfo {author} {\bibfnamefont {N.}~\bibnamefont {Yunes}},\ }\bibfield  {title} {\bibinfo {title} {Breaking bad degeneracies with love relations: {I}mproving gravitational-wave measurements through universal relations},\ }\href {https://doi.org/10.1103/PhysRevD.107.043010} {\bibfield  {journal} {\bibinfo  {journal} {Phys. Rev. D}\ }\textbf {\bibinfo {volume} {107}},\ \bibinfo {pages} {043010} (\bibinfo {year} {2023})}\BibitemShut {NoStop}%
\bibitem [{\citenamefont {Yüksel}\ \emph {et~al.}(2021)\citenamefont {Yüksel}, \citenamefont {Oishi},\ and\ \citenamefont {Paar}}]{Yuksel2021}%
  \BibitemOpen
  \bibfield  {author} {\bibinfo {author} {\bibfnamefont {E.}~\bibnamefont {Yüksel}}, \bibinfo {author} {\bibfnamefont {T.}~\bibnamefont {Oishi}},\ and\ \bibinfo {author} {\bibfnamefont {N.}~\bibnamefont {Paar}},\ }\bibfield  {title} {\bibinfo {title} {Nuclear {E}quation of {S}tate in the {R}elativistic {P}oint-{C}oupling {M}odel {C}onstrained by {E}xcitations in {F}inite {N}uclei},\ }\href {https://doi.org/10.3390/universe7030071} {\bibfield  {journal} {\bibinfo  {journal} {Universe}\ }\textbf {\bibinfo {volume} {7}},\ \bibinfo {pages} {71} (\bibinfo {year} {2021})}\BibitemShut {NoStop}%
\bibitem [{\citenamefont {Koliogiannis}\ \emph {et~al.}(2025{\natexlab{a}})\citenamefont {Koliogiannis}, \citenamefont {Yüksel},\ and\ \citenamefont {Paar}}]{KOLIOGIANNIS2025139362}%
  \BibitemOpen
  \bibfield  {author} {\bibinfo {author} {\bibfnamefont {P.}~\bibnamefont {Koliogiannis}}, \bibinfo {author} {\bibfnamefont {E.}~\bibnamefont {Yüksel}},\ and\ \bibinfo {author} {\bibfnamefont {N.}~\bibnamefont {Paar}},\ }\bibfield  {title} {\bibinfo {title} {Constraining neutron star properties through parity-violating electron scattering experiments and relativistic point coupling interactions},\ }\href {https://doi.org/https://doi.org/10.1016/j.physletb.2025.139362} {\bibfield  {journal} {\bibinfo  {journal} {Phys. Lett. B}\ }\textbf {\bibinfo {volume} {862}},\ \bibinfo {pages} {139362} (\bibinfo {year} {2025}{\natexlab{a}})}\BibitemShut {NoStop}%
\bibitem [{\citenamefont {Vretenar}\ \emph {et~al.}(2003)\citenamefont {Vretenar}, \citenamefont {Nik\ifmmode \check{s}\else \v{s}\fi{}i\ifmmode~\acute{c}\else \'{c}\fi{}},\ and\ \citenamefont {Ring}}]{PhysRevC.68.024310}%
  \BibitemOpen
  \bibfield  {author} {\bibinfo {author} {\bibfnamefont {D.}~\bibnamefont {Vretenar}}, \bibinfo {author} {\bibfnamefont {T.}~\bibnamefont {Nik\ifmmode \check{s}\else \v{s}\fi{}i\ifmmode~\acute{c}\else \'{c}\fi{}}},\ and\ \bibinfo {author} {\bibfnamefont {P.}~\bibnamefont {Ring}},\ }\bibfield  {title} {\bibinfo {title} {A microscopic estimate of the nuclear matter compressibility and symmetry energy in relativistic mean-field models},\ }\href {https://doi.org/10.1103/PhysRevC.68.024310} {\bibfield  {journal} {\bibinfo  {journal} {Phys. Rev. C}\ }\textbf {\bibinfo {volume} {68}},\ \bibinfo {pages} {024310} (\bibinfo {year} {2003})}\BibitemShut {NoStop}%
\bibitem [{\citenamefont {Reinhard}\ \emph {et~al.}(1986)\citenamefont {Reinhard}, \citenamefont {Rufa}, \citenamefont {Maruhn}, \citenamefont {Greiner},\ and\ \citenamefont {Friedrich}}]{Reinhard1986}%
  \BibitemOpen
  \bibfield  {author} {\bibinfo {author} {\bibfnamefont {P.-G.}\ \bibnamefont {Reinhard}}, \bibinfo {author} {\bibfnamefont {M.}~\bibnamefont {Rufa}}, \bibinfo {author} {\bibfnamefont {J.}~\bibnamefont {Maruhn}}, \bibinfo {author} {\bibfnamefont {W.}~\bibnamefont {Greiner}},\ and\ \bibinfo {author} {\bibfnamefont {J.}~\bibnamefont {Friedrich}},\ }\bibfield  {title} {\bibinfo {title} {Nuclear ground-state properties in a relativistic meson-field theory},\ }\href {https://doi.org/10.1007/BF01294551} {\bibfield  {journal} {\bibinfo  {journal} {Z. Phys. A}\ }\textbf {\bibinfo {volume} {323}},\ \bibinfo {pages} {13} (\bibinfo {year} {1986})}\BibitemShut {NoStop}%
\bibitem [{\citenamefont {Lalazissis}\ \emph {et~al.}(1997)\citenamefont {Lalazissis}, \citenamefont {K\"onig},\ and\ \citenamefont {Ring}}]{PhysRevC.55.540}%
  \BibitemOpen
  \bibfield  {author} {\bibinfo {author} {\bibfnamefont {G.~A.}\ \bibnamefont {Lalazissis}}, \bibinfo {author} {\bibfnamefont {J.}~\bibnamefont {K\"onig}},\ and\ \bibinfo {author} {\bibfnamefont {P.}~\bibnamefont {Ring}},\ }\bibfield  {title} {\bibinfo {title} {New parametrization for the lagrangian density of relativistic mean field theory},\ }\href {https://doi.org/10.1103/PhysRevC.55.540} {\bibfield  {journal} {\bibinfo  {journal} {Phys. Rev. C}\ }\textbf {\bibinfo {volume} {55}},\ \bibinfo {pages} {540} (\bibinfo {year} {1997})}\BibitemShut {NoStop}%
\bibitem [{\citenamefont {Sun}\ \emph {et~al.}(2024)\citenamefont {Sun}, \citenamefont {Bhattiprolu},\ and\ \citenamefont {Lattimer}}]{PhysRevC.109.055801}%
  \BibitemOpen
  \bibfield  {author} {\bibinfo {author} {\bibfnamefont {B.}~\bibnamefont {Sun}}, \bibinfo {author} {\bibfnamefont {S.}~\bibnamefont {Bhattiprolu}},\ and\ \bibinfo {author} {\bibfnamefont {J.~M.}\ \bibnamefont {Lattimer}},\ }\bibfield  {title} {\bibinfo {title} {Compiled properties of nucleonic matter and nuclear and neutron star models from nonrelativistic and relativistic interactions},\ }\href {https://doi.org/10.1103/PhysRevC.109.055801} {\bibfield  {journal} {\bibinfo  {journal} {Phys. Rev. C}\ }\textbf {\bibinfo {volume} {109}},\ \bibinfo {pages} {055801} (\bibinfo {year} {2024})}\BibitemShut {NoStop}%
\bibitem [{\citenamefont {Lalazissis}\ \emph {et~al.}(2005)\citenamefont {Lalazissis}, \citenamefont {Nik\ifmmode \check{s}\else \v{s}\fi{}i\ifmmode~\acute{c}\else \'{c}\fi{}}, \citenamefont {Vretenar},\ and\ \citenamefont {Ring}}]{PhysRevC.71.024312}%
  \BibitemOpen
  \bibfield  {author} {\bibinfo {author} {\bibfnamefont {G.~A.}\ \bibnamefont {Lalazissis}}, \bibinfo {author} {\bibfnamefont {T.}~\bibnamefont {Nik\ifmmode \check{s}\else \v{s}\fi{}i\ifmmode~\acute{c}\else \'{c}\fi{}}}, \bibinfo {author} {\bibfnamefont {D.}~\bibnamefont {Vretenar}},\ and\ \bibinfo {author} {\bibfnamefont {P.}~\bibnamefont {Ring}},\ }\bibfield  {title} {\bibinfo {title} {New relativistic mean-field interaction with density-dependent meson-nucleon couplings},\ }\href {https://doi.org/10.1103/PhysRevC.71.024312} {\bibfield  {journal} {\bibinfo  {journal} {Phys. Rev. C}\ }\textbf {\bibinfo {volume} {71}},\ \bibinfo {pages} {024312} (\bibinfo {year} {2005})}\BibitemShut {NoStop}%
\bibitem [{\citenamefont {Agrawal}\ \emph {et~al.}(2005)\citenamefont {Agrawal}, \citenamefont {Shlomo},\ and\ \citenamefont {Au}}]{PhysRevC.72.014310}%
  \BibitemOpen
  \bibfield  {author} {\bibinfo {author} {\bibfnamefont {B.~K.}\ \bibnamefont {Agrawal}}, \bibinfo {author} {\bibfnamefont {S.}~\bibnamefont {Shlomo}},\ and\ \bibinfo {author} {\bibfnamefont {V.~K.}\ \bibnamefont {Au}},\ }\bibfield  {title} {\bibinfo {title} {Determination of the parameters of a skyrme type effective interaction using the simulated annealing approach},\ }\href {https://doi.org/10.1103/PhysRevC.72.014310} {\bibfield  {journal} {\bibinfo  {journal} {Phys. Rev. C}\ }\textbf {\bibinfo {volume} {72}},\ \bibinfo {pages} {014310} (\bibinfo {year} {2005})}\BibitemShut {NoStop}%
\bibitem [{\citenamefont {{Van Giai}}\ and\ \citenamefont {Sagawa}(1981)}]{VANGIAI1981379}%
  \BibitemOpen
  \bibfield  {author} {\bibinfo {author} {\bibfnamefont {N.}~\bibnamefont {{Van Giai}}}\ and\ \bibinfo {author} {\bibfnamefont {H.}~\bibnamefont {Sagawa}},\ }\bibfield  {title} {\bibinfo {title} {Spin-isospin and pairing properties of modified skyrme interactions},\ }\href {https://doi.org/https://doi.org/10.1016/0370-2693(81)90646-8} {\bibfield  {journal} {\bibinfo  {journal} {Phys. Lett. B}\ }\textbf {\bibinfo {volume} {106}},\ \bibinfo {pages} {379} (\bibinfo {year} {1981})}\BibitemShut {NoStop}%
\bibitem [{\citenamefont {Agrawal}\ \emph {et~al.}(2003)\citenamefont {Agrawal}, \citenamefont {Shlomo},\ and\ \citenamefont {Kim~Au}}]{PhysRevC.68.031304}%
  \BibitemOpen
  \bibfield  {author} {\bibinfo {author} {\bibfnamefont {B.~K.}\ \bibnamefont {Agrawal}}, \bibinfo {author} {\bibfnamefont {S.}~\bibnamefont {Shlomo}},\ and\ \bibinfo {author} {\bibfnamefont {V.}~\bibnamefont {Kim~Au}},\ }\bibfield  {title} {\bibinfo {title} {Nuclear matter incompressibility coefficient in relativistic and nonrelativistic microscopic models},\ }\href {https://doi.org/10.1103/PhysRevC.68.031304} {\bibfield  {journal} {\bibinfo  {journal} {Phys. Rev. C}\ }\textbf {\bibinfo {volume} {68}},\ \bibinfo {pages} {031304} (\bibinfo {year} {2003})}\BibitemShut {NoStop}%
\bibitem [{\citenamefont {Chabanat}\ \emph {et~al.}(1997)\citenamefont {Chabanat}, \citenamefont {Bonche}, \citenamefont {Haensel}, \citenamefont {Meyer},\ and\ \citenamefont {Schaeffer}}]{CHABANAT1997710}%
  \BibitemOpen
  \bibfield  {author} {\bibinfo {author} {\bibfnamefont {E.}~\bibnamefont {Chabanat}}, \bibinfo {author} {\bibfnamefont {P.}~\bibnamefont {Bonche}}, \bibinfo {author} {\bibfnamefont {P.}~\bibnamefont {Haensel}}, \bibinfo {author} {\bibfnamefont {J.}~\bibnamefont {Meyer}},\ and\ \bibinfo {author} {\bibfnamefont {R.}~\bibnamefont {Schaeffer}},\ }\bibfield  {title} {\bibinfo {title} {A {S}kyrme parametrization from subnuclear to neutron star densities},\ }\href {https://doi.org/https://doi.org/10.1016/S0375-9474(97)00596-4} {\bibfield  {journal} {\bibinfo  {journal} {Nuc. Phys. A}\ }\textbf {\bibinfo {volume} {627}},\ \bibinfo {pages} {710} (\bibinfo {year} {1997})}\BibitemShut {NoStop}%
\bibitem [{\citenamefont {Chabanat}\ \emph {et~al.}(1998)\citenamefont {Chabanat}, \citenamefont {Bonche}, \citenamefont {Haensel}, \citenamefont {Meyer},\ and\ \citenamefont {Schaeffer}}]{CHABANAT1998231}%
  \BibitemOpen
  \bibfield  {author} {\bibinfo {author} {\bibfnamefont {E.}~\bibnamefont {Chabanat}}, \bibinfo {author} {\bibfnamefont {P.}~\bibnamefont {Bonche}}, \bibinfo {author} {\bibfnamefont {P.}~\bibnamefont {Haensel}}, \bibinfo {author} {\bibfnamefont {J.}~\bibnamefont {Meyer}},\ and\ \bibinfo {author} {\bibfnamefont {R.}~\bibnamefont {Schaeffer}},\ }\bibfield  {title} {\bibinfo {title} {A {S}kyrme parametrization from subnuclear to neutron star densities {P}art {II}. {N}uclei far from stabilities},\ }\href {https://doi.org/https://doi.org/10.1016/S0375-9474(98)00180-8} {\bibfield  {journal} {\bibinfo  {journal} {Nuc. Phys. A}\ }\textbf {\bibinfo {volume} {635}},\ \bibinfo {pages} {231} (\bibinfo {year} {1998})}\BibitemShut {NoStop}%
\bibitem [{\citenamefont {Reinhard}\ and\ \citenamefont {Flocard}(1995)}]{REINHARD1995467}%
  \BibitemOpen
  \bibfield  {author} {\bibinfo {author} {\bibfnamefont {P.-G.}\ \bibnamefont {Reinhard}}\ and\ \bibinfo {author} {\bibfnamefont {H.}~\bibnamefont {Flocard}},\ }\bibfield  {title} {\bibinfo {title} {Nuclear effective forces and isotope shifts},\ }\href {https://doi.org/https://doi.org/10.1016/0375-9474(94)00770-N} {\bibfield  {journal} {\bibinfo  {journal} {Nuclear Physics A}\ }\textbf {\bibinfo {volume} {584}},\ \bibinfo {pages} {467} (\bibinfo {year} {1995})}\BibitemShut {NoStop}%
\bibitem [{\citenamefont {Roca-Maza}\ \emph {et~al.}(2012)\citenamefont {Roca-Maza}, \citenamefont {Col\`o},\ and\ \citenamefont {Sagawa}}]{PhysRevC.86.031306}%
  \BibitemOpen
  \bibfield  {author} {\bibinfo {author} {\bibfnamefont {X.}~\bibnamefont {Roca-Maza}}, \bibinfo {author} {\bibfnamefont {G.}~\bibnamefont {Col\`o}},\ and\ \bibinfo {author} {\bibfnamefont {H.}~\bibnamefont {Sagawa}},\ }\bibfield  {title} {\bibinfo {title} {New skyrme interaction with improved spin-isospin properties},\ }\href {https://doi.org/10.1103/PhysRevC.86.031306} {\bibfield  {journal} {\bibinfo  {journal} {Phys. Rev. C}\ }\textbf {\bibinfo {volume} {86}},\ \bibinfo {pages} {031306} (\bibinfo {year} {2012})}\BibitemShut {NoStop}%
\bibitem [{\citenamefont {Douchin}\ and\ \citenamefont {Haensel}(2001)}]{Douchin_2001}%
  \BibitemOpen
  \bibfield  {author} {\bibinfo {author} {\bibfnamefont {F.}~\bibnamefont {Douchin}}\ and\ \bibinfo {author} {\bibfnamefont {P.}~\bibnamefont {Haensel}},\ }\bibfield  {title} {\bibinfo {title} {A unified equation of state of dense matter and neutron star structure},\ }\href {https://doi.org/10.1051/0004-6361:20011402} {\bibfield  {journal} {\bibinfo  {journal} {Astron. Astrophys.}\ }\textbf {\bibinfo {volume} {380}},\ \bibinfo {pages} {151} (\bibinfo {year} {2001})}\BibitemShut {NoStop}%
\bibitem [{\citenamefont {{Baym}}\ \emph {et~al.}(1971)\citenamefont {{Baym}}, \citenamefont {{Pethick}},\ and\ \citenamefont {{Sutherland}}}]{Baym-71}%
  \BibitemOpen
  \bibfield  {author} {\bibinfo {author} {\bibfnamefont {G.}~\bibnamefont {{Baym}}}, \bibinfo {author} {\bibfnamefont {C.}~\bibnamefont {{Pethick}}},\ and\ \bibinfo {author} {\bibfnamefont {P.}~\bibnamefont {{Sutherland}}},\ }\bibfield  {title} {\bibinfo {title} {{{T}he {G}round {S}tate of {M}atter at {H}igh {D}ensities: {E}quation of {S}tate and {S}tellar {M}odels}},\ }\href {https://doi.org/10.1086/151216} {\bibfield  {journal} {\bibinfo  {journal} {Astrophys. J.}\ }\textbf {\bibinfo {volume} {170}},\ \bibinfo {pages} {299} (\bibinfo {year} {1971})}\BibitemShut {NoStop}%
\bibitem [{\citenamefont {Feynman}\ \emph {et~al.}(1949)\citenamefont {Feynman}, \citenamefont {Metropolis},\ and\ \citenamefont {Teller}}]{PhysRev.75.1561}%
  \BibitemOpen
  \bibfield  {author} {\bibinfo {author} {\bibfnamefont {R.~P.}\ \bibnamefont {Feynman}}, \bibinfo {author} {\bibfnamefont {N.}~\bibnamefont {Metropolis}},\ and\ \bibinfo {author} {\bibfnamefont {E.}~\bibnamefont {Teller}},\ }\bibfield  {title} {\bibinfo {title} {Equations of {S}tate of {E}lements {B}ased on the {G}eneralized {F}ermi-{T}homas {T}heory},\ }\href {https://doi.org/10.1103/PhysRev.75.1561} {\bibfield  {journal} {\bibinfo  {journal} {Phys. Rev.}\ }\textbf {\bibinfo {volume} {75}},\ \bibinfo {pages} {1561} (\bibinfo {year} {1949})}\BibitemShut {NoStop}%
\bibitem [{\citenamefont {Cai}\ and\ \citenamefont {Chen}(2012)}]{PhysRevC.85.024302}%
  \BibitemOpen
  \bibfield  {author} {\bibinfo {author} {\bibfnamefont {B.-J.}\ \bibnamefont {Cai}}\ and\ \bibinfo {author} {\bibfnamefont {L.-W.}\ \bibnamefont {Chen}},\ }\bibfield  {title} {\bibinfo {title} {Nuclear matter fourth-order symmetry energy in the relativistic mean field models},\ }\href {https://doi.org/10.1103/PhysRevC.85.024302} {\bibfield  {journal} {\bibinfo  {journal} {Phys. Rev. C}\ }\textbf {\bibinfo {volume} {85}},\ \bibinfo {pages} {024302} (\bibinfo {year} {2012})}\BibitemShut {NoStop}%
\bibitem [{\citenamefont {Paar}\ \emph {et~al.}(2014)\citenamefont {Paar}, \citenamefont {Moustakidis}, \citenamefont {Marketin}, \citenamefont {Vretenar},\ and\ \citenamefont {Lalazissis}}]{PhysRevC.90.011304}%
  \BibitemOpen
  \bibfield  {author} {\bibinfo {author} {\bibfnamefont {N.}~\bibnamefont {Paar}}, \bibinfo {author} {\bibfnamefont {C.~C.}\ \bibnamefont {Moustakidis}}, \bibinfo {author} {\bibfnamefont {T.}~\bibnamefont {Marketin}}, \bibinfo {author} {\bibfnamefont {D.}~\bibnamefont {Vretenar}},\ and\ \bibinfo {author} {\bibfnamefont {G.~A.}\ \bibnamefont {Lalazissis}},\ }\bibfield  {title} {\bibinfo {title} {Neutron star structure and collective excitations of finite nuclei},\ }\href {https://doi.org/10.1103/PhysRevC.90.011304} {\bibfield  {journal} {\bibinfo  {journal} {Phys. Rev. C}\ }\textbf {\bibinfo {volume} {90}},\ \bibinfo {pages} {011304} (\bibinfo {year} {2014})}\BibitemShut {NoStop}%
\bibitem [{\citenamefont {Doroshenko}\ \emph {et~al.}(2022)\citenamefont {Doroshenko}, \citenamefont {Suleimanov}, \citenamefont {Pühlhofer},\ and\ \citenamefont {Santangelo}}]{Doroshenko-2022}%
  \BibitemOpen
  \bibfield  {author} {\bibinfo {author} {\bibfnamefont {V.}~\bibnamefont {Doroshenko}}, \bibinfo {author} {\bibfnamefont {V.}~\bibnamefont {Suleimanov}}, \bibinfo {author} {\bibfnamefont {G.}~\bibnamefont {Pühlhofer}},\ and\ \bibinfo {author} {\bibfnamefont {A.}~\bibnamefont {Santangelo}},\ }\bibfield  {title} {\bibinfo {title} {A strangely light neutron star within a supernova remnant.},\ }\href {https://doi.org/https://doi.org/10.1038/s41550-022-01800-1} {\bibfield  {journal} {\bibinfo  {journal} {Nat. Astron.}\ }\textbf {\bibinfo {volume} {6}},\ \bibinfo {pages} {1444–1451} (\bibinfo {year} {2022})}\BibitemShut {NoStop}%
\bibitem [{\citenamefont {Choudhury}\ \emph {et~al.}(2024)\citenamefont {Choudhury} \emph {et~al.}}]{Choudhury_2024}%
  \BibitemOpen
  \bibfield  {author} {\bibinfo {author} {\bibfnamefont {D.}~\bibnamefont {Choudhury}} \emph {et~al.},\ }\bibfield  {title} {\bibinfo {title} {A {NICER} {V}iew of the {N}earest and {B}rightest {M}illisecond {P}ulsar: {PSR} {J}0437–4715},\ }\href {https://doi.org/10.3847/2041-8213/ad5a6f} {\bibfield  {journal} {\bibinfo  {journal} {Astrophys. J. Lett.}\ }\textbf {\bibinfo {volume} {971}},\ \bibinfo {pages} {L20} (\bibinfo {year} {2024})}\BibitemShut {NoStop}%
\bibitem [{\citenamefont {Salmi}\ \emph {et~al.}(2024)\citenamefont {Salmi} \emph {et~al.}}]{Salmi_2024}%
  \BibitemOpen
  \bibfield  {author} {\bibinfo {author} {\bibfnamefont {T.}~\bibnamefont {Salmi}} \emph {et~al.},\ }\bibfield  {title} {\bibinfo {title} {A {NICER} {V}iew of {PSR} {J}1231-1411: {A} {C}omplex {C}ase},\ }\href {https://doi.org/10.3847/1538-4357/ad81d2} {\bibfield  {journal} {\bibinfo  {journal} {Astrophys. J.}\ }\textbf {\bibinfo {volume} {976}},\ \bibinfo {pages} {58} (\bibinfo {year} {2024})}\BibitemShut {NoStop}%
\bibitem [{\citenamefont {Koliogiannis}\ \emph {et~al.}(2025{\natexlab{b}})\citenamefont {Koliogiannis}, \citenamefont {Yüksel}, \citenamefont {Ghosh},\ and\ \citenamefont {Paar}}]{Koliogiannis_2026}%
  \BibitemOpen
  \bibfield  {author} {\bibinfo {author} {\bibfnamefont {P.}~\bibnamefont {Koliogiannis}}, \bibinfo {author} {\bibfnamefont {E.}~\bibnamefont {Yüksel}}, \bibinfo {author} {\bibfnamefont {T.}~\bibnamefont {Ghosh}},\ and\ \bibinfo {author} {\bibfnamefont {N.}~\bibnamefont {Paar}},\ }\bibfield  {title} {\bibinfo {title} {Dipole {P}olarizability of {F}inite {N}uclei as a {P}robe of {N}eutron {S}tars},\ }\href {https://doi.org/10.3847/2041-8213/ae261d} {\bibfield  {journal} {\bibinfo  {journal} {Astrophys. J. Lett.}\ }\textbf {\bibinfo {volume} {996}},\ \bibinfo {pages} {L18} (\bibinfo {year} {2025}{\natexlab{b}})}\BibitemShut {NoStop}%
\bibitem [{\citenamefont {Paar}\ \emph {et~al.}(2007)\citenamefont {Paar}, \citenamefont {Vretenar}, \citenamefont {Khan},\ and\ \citenamefont {Colò}}]{Paar_2007}%
  \BibitemOpen
  \bibfield  {author} {\bibinfo {author} {\bibfnamefont {N.}~\bibnamefont {Paar}}, \bibinfo {author} {\bibfnamefont {D.}~\bibnamefont {Vretenar}}, \bibinfo {author} {\bibfnamefont {E.}~\bibnamefont {Khan}},\ and\ \bibinfo {author} {\bibfnamefont {G.}~\bibnamefont {Colò}},\ }\bibfield  {title} {\bibinfo {title} {Exotic modes of excitation in atomic nuclei far from stability},\ }\href {https://doi.org/10.1088/0034-4885/70/5/R02} {\bibfield  {journal} {\bibinfo  {journal} {Rep. Prog. Phys.}\ }\textbf {\bibinfo {volume} {70}},\ \bibinfo {pages} {R02} (\bibinfo {year} {2007})}\BibitemShut {NoStop}%
\bibitem [{\citenamefont {Roca-Maza}\ \emph {et~al.}(2013)\citenamefont {Roca-Maza}, \citenamefont {Brenna}, \citenamefont {Col\`o}, \citenamefont {Centelles}, \citenamefont {Vi\~nas}, \citenamefont {Agrawal}, \citenamefont {Paar}, \citenamefont {Vretenar},\ and\ \citenamefont {Piekarewicz}}]{Rocamaza2013}%
  \BibitemOpen
  \bibfield  {author} {\bibinfo {author} {\bibfnamefont {X.}~\bibnamefont {Roca-Maza}}, \bibinfo {author} {\bibfnamefont {M.}~\bibnamefont {Brenna}}, \bibinfo {author} {\bibfnamefont {G.}~\bibnamefont {Col\`o}}, \bibinfo {author} {\bibfnamefont {M.}~\bibnamefont {Centelles}}, \bibinfo {author} {\bibfnamefont {X.}~\bibnamefont {Vi\~nas}}, \bibinfo {author} {\bibfnamefont {B.~K.}\ \bibnamefont {Agrawal}}, \bibinfo {author} {\bibfnamefont {N.}~\bibnamefont {Paar}}, \bibinfo {author} {\bibfnamefont {D.}~\bibnamefont {Vretenar}},\ and\ \bibinfo {author} {\bibfnamefont {J.}~\bibnamefont {Piekarewicz}},\ }\bibfield  {title} {\bibinfo {title} {Electric dipole polarizability in $^{208}\mathrm{Pb}$: Insights from the droplet model},\ }\href {https://doi.org/10.1103/PhysRevC.88.024316} {\bibfield  {journal} {\bibinfo  {journal} {Phys. Rev. C}\ }\textbf {\bibinfo {volume} {88}},\ \bibinfo {pages} {024316} (\bibinfo {year} {2013})}\BibitemShut {NoStop}%
\bibitem [{\citenamefont {Paar}\ \emph {et~al.}(2003)\citenamefont {Paar}, \citenamefont {Ring}, \citenamefont {Nik\ifmmode \check{s}\else \v{s}\fi{}i\ifmmode~\acute{c}\else \'{c}\fi{}},\ and\ \citenamefont {Vretenar}}]{PhysRevC.67.034312}%
  \BibitemOpen
  \bibfield  {author} {\bibinfo {author} {\bibfnamefont {N.}~\bibnamefont {Paar}}, \bibinfo {author} {\bibfnamefont {P.}~\bibnamefont {Ring}}, \bibinfo {author} {\bibfnamefont {T.}~\bibnamefont {Nik\ifmmode \check{s}\else \v{s}\fi{}i\ifmmode~\acute{c}\else \'{c}\fi{}}},\ and\ \bibinfo {author} {\bibfnamefont {D.}~\bibnamefont {Vretenar}},\ }\bibfield  {title} {\bibinfo {title} {Quasiparticle random phase approximation based on the relativistic {H}artree-{B}ogoliubov model},\ }\href {https://doi.org/10.1103/PhysRevC.67.034312} {\bibfield  {journal} {\bibinfo  {journal} {Phys. Rev. C}\ }\textbf {\bibinfo {volume} {67}},\ \bibinfo {pages} {034312} (\bibinfo {year} {2003})}\BibitemShut {NoStop}%
\bibitem [{\citenamefont {Colò}\ \emph {et~al.}(2013)\citenamefont {Colò}, \citenamefont {Cao}, \citenamefont {{Van Giai}},\ and\ \citenamefont {Capelli}}]{colo2013self}%
  \BibitemOpen
  \bibfield  {author} {\bibinfo {author} {\bibfnamefont {G.}~\bibnamefont {Colò}}, \bibinfo {author} {\bibfnamefont {L.}~\bibnamefont {Cao}}, \bibinfo {author} {\bibfnamefont {N.}~\bibnamefont {{Van Giai}}},\ and\ \bibinfo {author} {\bibfnamefont {L.}~\bibnamefont {Capelli}},\ }\bibfield  {title} {\bibinfo {title} {Self-consistent {RPA} calculations with {S}kyrme-type interactions: The skyrme\_rpa program},\ }\href {https://doi.org/10.1016/j.cpc.2012.07.016} {\bibfield  {journal} {\bibinfo  {journal} {Comp. Phys. Comm.}\ }\textbf {\bibinfo {volume} {184}},\ \bibinfo {pages} {142} (\bibinfo {year} {2013})}\BibitemShut {NoStop}%
\bibitem [{\citenamefont {Colò}\ and\ \citenamefont {Roca-Maza}(2021)}]{colo2021userguidehfbcsqrpav1code}%
  \BibitemOpen
  \bibfield  {author} {\bibinfo {author} {\bibfnamefont {G.}~\bibnamefont {Colò}}\ and\ \bibinfo {author} {\bibfnamefont {X.}~\bibnamefont {Roca-Maza}},\ }\href {https://arxiv.org/abs/2102.06562} {\bibinfo {title} {User guide for the hfbcs-qrpa(v1) code}} (\bibinfo {year} {2021}),\ \Eprint {https://arxiv.org/abs/2102.06562} {arXiv:2102.06562 [nucl-th]} \BibitemShut {NoStop}%
\bibitem [{\citenamefont {Reinhard}\ and\ \citenamefont {Nazarewicz}(2010)}]{PhysRevC.81.051303}%
  \BibitemOpen
  \bibfield  {author} {\bibinfo {author} {\bibfnamefont {P.-G.}\ \bibnamefont {Reinhard}}\ and\ \bibinfo {author} {\bibfnamefont {W.}~\bibnamefont {Nazarewicz}},\ }\bibfield  {title} {\bibinfo {title} {Information content of a new observable: The case of the nuclear neutron skin},\ }\href {https://doi.org/10.1103/PhysRevC.81.051303} {\bibfield  {journal} {\bibinfo  {journal} {Phys. Rev. C}\ }\textbf {\bibinfo {volume} {81}},\ \bibinfo {pages} {051303} (\bibinfo {year} {2010})}\BibitemShut {NoStop}%
\bibitem [{\citenamefont {Capano}\ \emph {et~al.}(2020)\citenamefont {Capano} \emph {et~al.}}]{Capano-2020}%
  \BibitemOpen
  \bibfield  {author} {\bibinfo {author} {\bibfnamefont {C.~D.}\ \bibnamefont {Capano}} \emph {et~al.},\ }\bibfield  {title} {\bibinfo {title} {Stringent constraints on neutron-star radii from multimessenger observations and nuclear theory},\ }\href {https://doi.org/10.1038/s41550-020-1014-6} {\bibfield  {journal} {\bibinfo  {journal} {Nat. Astron.}\ }\textbf {\bibinfo {volume} {4}},\ \bibinfo {pages} {625} (\bibinfo {year} {2020})}\BibitemShut {NoStop}%
\bibitem [{\citenamefont {Dietrich}\ \emph {et~al.}(2020)\citenamefont {Dietrich} \emph {et~al.}}]{doi:10.1126/science.abb4317}%
  \BibitemOpen
  \bibfield  {author} {\bibinfo {author} {\bibfnamefont {T.}~\bibnamefont {Dietrich}} \emph {et~al.},\ }\bibfield  {title} {\bibinfo {title} {Multimessenger constraints on the neutron-star equation of state and the {H}ubble constant},\ }\href {https://doi.org/10.1126/science.abb4317} {\bibfield  {journal} {\bibinfo  {journal} {Sci.}\ }\textbf {\bibinfo {volume} {370}},\ \bibinfo {pages} {1450} (\bibinfo {year} {2020})}\BibitemShut {NoStop}%
\end{thebibliography}%
%======================================================================================================================
\end{document}